\documentclass[useAMS,usenatbib,referee]{biom}

\usepackage{amsmath}
\usepackage{amsfonts}
\usepackage{graphicx}
\usepackage{epstopdf}
\usepackage{color}
\usepackage{breqn}
\usepackage{calrsfs}
\usepackage{subcaption}

\usepackage{bbm}

\newtheorem{proposition}[theorem]{Proposition}

\usepackage{tikz-cd}
\usetikzlibrary{fit,shapes.geometric}
\tikzset{mynode/.style={
    rectangle,draw,
    inner xsep=-1pt, black,
    }}
\usetikzlibrary{calc}

\DeclareMathAlphabet{\pazocal}{OMS}{zplm}{m}{n}

\newcommand{\Ib}{\pazocal{I}}

 \newcommand{\bitem}{\begin{itemize}}
 \newcommand{\eitem}{\end{itemize}}

\usepackage[figuresright]{rotating}

\title[A general modelling framework for open wildlife populations based on the Polya Tree prior]{A general modelling framework for open wildlife populations based on the Polya Tree prior}

  \author{\hspace{-1cm} Alex Diana \hspace{-1cm}\\
    School of Mathematics, Statistics and Actuarial Science, University of Kent, UK\\    
    Eleni Matechou \\
    School of Mathematics, Statistics and Actuarial Science, University of Kent, UK \\
    \hspace{-1cm} Jim Griffin \\
    Department of Statistical Science, University College London, UK \\
     \hspace{-1cm} Todd Arnold \\
    \hspace{-1.5cm} Department of Fisheries, Wildlife and Conservation Biology, University of Minnesota, USA \\
    \hspace{-1cm} Richard A. Griffiths \\
    \hspace{0cm} Durrell Institute of Conservation and Ecology, \\ School of Anthropology and Conservation, University of Kent, UK \\
    \hspace{-1.5cm} John Pickering \\
    University of Georgia, Athens, USA \\
    \hspace{-1cm} Simone Tenan \\
    Institute of Marine Sciences (CNR-ISMAR), Venezia, Italy \\
    \hspace{-1cm} Stefano Volponi \\
    Istituto Superiore Protezione e Ricerca Ambientale, Bologna, Italy 
}

\begin{document}



\pagerange{\pageref{firstpage}--\pageref{lastpage}} 


\label{firstpage}


\begin{abstract}
Wildlife monitoring for open populations can be performed using a number of different survey methods. Each survey method gives rise to a type of data and, in the last five decades, a large number of associated statistical models have been developed for analysing these data. Although these models have been parameterised and fitted using different approaches, they have all been designed to model the pattern with which individuals enter and exit the population and to estimate the population size. However, existing approaches rely on a predefined model structure and complexity, either by assuming that parameters are specific to sampling occasions, or by employing parametric curves. Instead, we propose a novel Bayesian nonparametric framework for modelling entry and exit patterns based on the Polya Tree (PT) prior for densities. Our Bayesian non-parametric approach avoids overfitting when inferring entry and exit patterns while simultaneously allowing more flexibility than is possible using parametric curves. We apply our new framework to capture-recapture, count and ring-recovery data and we introduce the replicated PT prior for defining classes of models for these data. Additionally, we define the Hierarchical Logistic PT prior for jointly modelling related data and we consider the Optional PT prior for modelling long time series of data. We demonstrate our new approach using five different case studies on birds, amphibians and insects.
\end{abstract}


\maketitle

\section{Introduction}


In recent years, there has been an increased interest in monitoring wildlife populations due to the ongoing effects of climate change and habitat destruction. However, monitoring wildlife populations is challenging because it is difficult to observe all animals present in the wild. Therefore, statistical models need to be employed to infer population sizes \citep{royle2004n}, migration \citep{pledger2009stopover}, phenological \citep{dennis2017using} or survival patterns \citep{mccrea2013age} from ecological data at one or more sites. 

Although these models are developed for data collected using different sampling protocols, they all focus on the estimation of the entry, exit and length of stay (LOS) patterns of populations, where entry can correspond to arrival/birth, exit to departure/death and LOS to retention/survival at the site. These existing methods either assume that parameters related to entry and exit patterns are completely unconstrained, requiring two parameters to be estimated for each sampling occasion, one for entry and one for exit \citep{pledger2009stopover,lyons2016population}, or use parametric curves to constrain entry \citep{matechou2014monitoring} and exit patterns \citep{mccrea2013age, jimenez2019estimating}.  In recent years, Bayesian nonparametric (BNP) methods have been used in ecology to flexibly model entry and exit distributions without making parametric assumptions. In particular, models have been built using the Dirichlet Process (DP) mixture prior \citep{lo1984class}, which is a model for continuous distributions relying on a mixture model, see for example \citet{ford2015modelling, matechou2017modelling, diana2020hierarchical}, or using the Polya Tree (PT) prior \citep{ferguson1974prior}, such as in \citet{diana2018polya}.



In ecological surveys, data are typically collected at discrete times, referred to as sampling occasions. Every pair of consecutive sampling occasions defines an interval, during which individuals may enter or exit the study area. No data are available for the periods between sampling occasions. This suggests that when inferring entry and exit patterns we need to model the grid defined by the sampling occasions in the bivariate space of entry and exit. In this paper, we develop a general BNP framework for modelling this grid, based on the PT prior, which was also considered by Diana et al. (2018). The PT prior is a nonparametric prior for densities that is constructed by defining an infinite sequence of nested partitions of the sample space and recursively assigning the masses of the elements of each partition. In our case, we only need to define the PT for a finite sequence of partitions up to the level of the grid, which defines a prior distribution for the probabilities of an individual to fall in each cell of the grid. Our new modelling framework allows us to build models directly on the distribution of entry, exit and LOS, with minimal parametric assumptions on the shape of these distributions. We note that, in contrast to existing BNP approaches based on DP mixtures that build continuous models, our approach allows us to make inference directly from the latent number of individuals in each cell of the grid, with computational complexity only depending on the number of sampling occasions and not the number of individuals in the population, leading to substantial computational advantages.

Another advantage of the PT framework is that it can be extended in various directions. To extend the model to data collected at different sites or years, we define the replicate PT (RPT), which shares random variables (r.v.s) across different PTs, allowing us to impose constraints on model parameters, and the Hierarchical Logistic PT (HLPT), which allows us to model r.v.s across different distributions in a hierarchical fashion, implying exchangeability across the different units. 

We apply our approach to four different case studies. The first considers the estimation of entry and exit densities by a joint model of capture-recapture (CR) and count data (CD) collected on the same population. The second considers the estimation of age-specific survival, i.e. LOS, using ring-recovery (RR) data, where only some of the individuals are of known age, and demonstrates the advantage of the PT in modelling survival from sparse data, without resorting to the use of parametric curves. The third application involves estimating entry and exit densities at different sites using the HLPT. Finally, the fourth scenario considers a long time-series of CD. For long time-series, choosing a partition at the frequency that the data are sampled leads to a large number of parameters. To overcome this problem, we use the Optional Polya Tree (OPT) of Wong and Ma (2010), which is an extension of the PT and allows us to infer the partition structure.

The paper is structured in the following way. In Section \ref{sec:sampling} we provide an overview of the sampling schemes considered as case studies in the paper. In Section \ref{PTsection} we describe the main features of the PT prior. In Section \ref{sec:unip} we present and motivate the univariate partitions of the PT used in the paper and demonstrate the concept using CR data. The RPT prior is defined in Section \ref{sec:rpt}. In Section \ref{sec:bivp} we define the bivariate partitions of the PT and define the joint model for CR data and CD and the model for RR data. The HLPT is defined in Section \ref{sec:hierarchical} and the use of the OPT is described in Section \ref{sec:moths}, with both sections presenting an application to CD collected across several species and years, respectively. Section \ref{sec:concl} concludes the paper and introduces some potential future directions.

\section{Background}

\subsection{Sampling schemes of population surveys}
\label{sec:sampling}

In this paper, we consider models for data resulting from three types of ecological sampling schemes: CR, CD and RR. In all three cases, we assume sampling is performed on $K$ sampling occasions at times $t_1,\dots,t_K$. 

In the case of CR data, on each sampling occasion, a sample of the individuals is caught, and newly caught individuals are uniquely marked. If $D$ is the total number of captured individuals, the captures can be summarised in a $D \times K$ matrix, $H$, where $H_{ik}$ is $1$ if the $i$-th individual was captured on sampling occasion $k$ and $0$ otherwise. Each row of the matrix is called a capture history. We define instead as \textit{presence history} the vector with entries equal to $1$ on sampling occasions when the individual was available for capture and $0$ otherwise. A commonly employed approach for modelling CR data is the Cormack-Jolly-Seber (CJS) model \citep{cormack1964estimates, jolly1965explicit, seber1965note}. The CJS model is built conditional on the time of first capture of each individual. After the first capture, alive individuals of age $a$ at time $t_j$ are assumed to remain alive at the site until the following sampling occasion at time $t_{j+1}$ with probability $\phi_{j,a}$ and, conditionally on being present at the site, they can be recaptured (or resighted) with probability $p$.  Since the CJS model conditions on first capture, it does not allow the estimation of the population size but only of $\phi_{j,a}$ and $p$.  In the following, we will always use the indexes $f$, $k$ and $l$ to indicate the times of first possible capture, first capture and last possible capture, respectively. We express the CJS model in a PT framework in Section \ref{sec:cjs}. 

CD are collected by visiting the site and simply detecting a number of individuals, without attempting to individually identify them. The data can be summarised in a vector $(C_1,\dots,C_K)$,  where $C_j$ denotes the number of individuals detected on sampling occasion $j$.  


Finally, according to the RR protocol, individuals are captured and marked each year and, after being marked, they can be recovered soon after they die with probability $\lambda$. The data are usually summarised in an upper-triangular $K \times K$ matrix, $R$, where cell $(i,j)$ denotes the number of individuals marked in year $i$ and recovered dead in year $j$, and in a vector $(m_1,\dots,m_K)$ where $m_k$ denotes the total number of individuals marked in year $k$.



\subsection{The Polya Tree prior}
\label{PTsection}

The Polya tree (PT) prior \citep{ferguson1974prior} is a nonparametric prior for probability distributions. It has two parameter sets: the first is a sequence of nested partitions $\Pi$ of the sample space $\Omega$, while the second parameter, $\alpha$, is a sequence of positive numbers associated with each set of each partition. 

The partition at the first level, $\pi_1$, is obtained by splitting the sample space in $K_{\Omega}$ sets $\{ B_{0},\dots,B_{K_{\Omega} - 1}\}$. Subsequently, to build the partition at the second level, $\pi_2$, each set $B_{i}$ is split into $K_i$ sets $\{B_{i0},\dots,B_{i K_i - 1}\}$, next each set $B_{ij}$ is split into $K_{ij}$ sets $\{ B_{ij0}, \dots, B_{ij K_{ij} - 1} \}$ and so on. We note that for dyadic splits, the Dirichlet distributions are replaced by Beta distributions.

By defining $\epsilon_1 \dots \epsilon_m$ as a generic sequence of positive integers (such that $\epsilon_i \in \{0,\dots,K_{\epsilon_1 \dots \epsilon_{i-1}} - 1 \}$) and $B_{\epsilon_1 \dots \epsilon_m}$ as the correspondent set of the partition, the random mass associated to $B_{\epsilon_1 \dots \epsilon_m}$ by the PT is given by $G(B_{\epsilon_1 \dots \epsilon_m}) = \prod_{i = 1}^m V_{\epsilon_1 \dots \epsilon_i}$, where $( V_{0}, \dots, V_{K_{\Omega}-1} ) \sim \text{Dirichlet}(\alpha_{0}, \dots, \alpha_{K_{\Omega}-1})$ and $( V_{\epsilon_1 \dots \epsilon_{i-1}0}, \dots, V_{\epsilon_1 \dots \epsilon_{i-1} K_{\epsilon_1 \dots \epsilon_{i-1}}-1} ) \sim \text{Dirichlet}(\alpha_{\epsilon_1 \dots \epsilon_{i-1}0}, \dots, \alpha_{\epsilon_1 \dots \epsilon_{i-1} K_{\epsilon_1 \dots \epsilon_{i-1}}-1})$ for $i > 1$. For example, in the case where splits are triadic, $G(B_{02})=V_0 V_{02}$ where $(V_0,V_1,V_2) \sim$ Dirichlet$(\alpha_0, \alpha_1, \alpha_2)$ and $(V_{00},V_{01}, V_{02}) \sim$ Dirichlet$(\alpha_{00}, \alpha_{01}, \alpha_{02})$. The $V$s can be interpreted as the conditional probabilities of falling in a particular set in the next level of the tree, that is, given $y \sim G$, where $G \sim \text{PT}$, $\mathbb{P}(y \in B_{\epsilon_1 \dots \epsilon_m} |y \in B_{\epsilon_1 \dots \epsilon_{m-1}} ) = V_{\epsilon_1\dots\epsilon_m}$.


The PT is a conjugate prior for a distribution $G$ if we have observations $y_1,\dots,y_n \stackrel{\text{i.i.d.}}{\sim} G$, since the posterior distribution $G \ | \ y_1,\dots,y_n$ is a PT with parameters $\Pi$ and $\alpha^{\star}$, where $\alpha^{\star}_{\epsilon} = \alpha_{\epsilon} + n_{\epsilon}$ and $n_{\epsilon}$ is the number of observations falling into set $B_{\epsilon}$. The PT can be centered on any distribution $G_0$ in the sense that, for every set $B$ of the partition, $\mathbb{E}[G(B)] = G_0(B)$. 


\section{Univariate PT models}
\label{sec:unip}

\begin{figure}[h]
\begin{tabular}{lll}
\includegraphics[width=12cm]{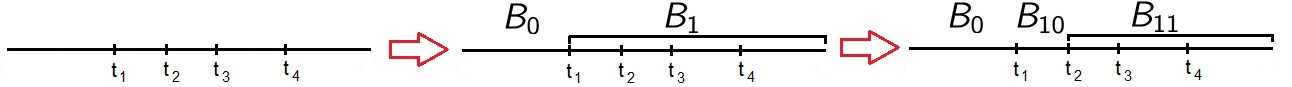}
 & 
 \begin{tikzpicture}[baseline= (a).base]
\node[scale=1] (a) at (0,0){
\begin{tikzcd}[column sep = -10pt]
& \Omega \arrow[ld, no head, swap, "V_0"] \arrow[rd, no head, ""]    & & & &  \\
B_0  & & B_1 \arrow[ld, no head, swap, "V_1"] \arrow[rd, no head, ""] & & & & &  \\
&  B_{10} & &  B_{11} \arrow[ld, no head, swap, "V_2"] \arrow[rd, no head, ""]  &  &  \\
& &     B_{110} & &  B_{111}  & 
\end{tikzcd}
};
\end{tikzpicture}
\end{tabular}
\caption{Sequence of splits (left) and structure (right) of the L94 partition. The r.v. assigning the ratio of the masses of two sets for each branch is represented on the left branch. The mass assigned to the set in each right branch is always one minus the mass assigned to the set in the corresponding left branch.}
\label{fig:L94}
\end{figure}

The partition of the PT can be chosen according to the specific application. To define a prior for the distribution of a time-to-event on the real line, \citet{lavine1994more} (L94) proposed the following partition. Assume we have sampling times $t_1,\dots,t_K \in \mathbb{R}$ and we only observe the interval where each time-to-event $x$ happens, that is, we observe whether $x < t_1$, $t_1 < x < t_2$ and so on. L94 proposed to define $\Pi$ such that $B_0 = (-\infty, t_1)$, $B_1 = (t_1, \infty)$, with $B_1$ further split into $B_{10} = (t_1, t_2)$, $B_{11} = (t_2, \infty)$ and so on. A representation is shown in Fig. \ref{fig:L94}. Following the notation of Section \ref{PTsection}, this structure is obtained when $K_{B_0} = K_{B_{10}} = K_{B_{110}} = \dots = 1$ and $K_{\Omega} = K_{B_1} = K_{B_{11}} = K_{B_{111}} = \dots = 2$ .

Samples from a PT with the partition defined above have decreasing variance starting from the earlier sampling times. This follows easily from the following proposition, proven for simplicity for Beta r.v.s with mean $1/2$.\begin{proposition}
Given $X_1 \sim \rm{Beta}$$(\alpha, \alpha)$ and $X_2 \sim \rm{Beta}$$(\beta, \beta)$, if $\beta \ge \alpha$ then $\rm{Var}(X_1 X_2) \le  \rm{Var}(X_1)$.
\end{proposition}
The previous proposition shows that if the r.v.s $V_0, V_1$, $\dots$ assigning the masses to the sets $B_0$, $B_1$, $\dots$ have parameters whose sum is not decreasing, the variance of $V_0$, $V_0 V_1$, $V_0 V_1 V_2$, $\dots$ is decreasing. Conversely, we can assume that the variance is decreasing starting from the last sampling times by reversing the order of the split and splitting the partition using first $t_K$, then $t_{K-1}$ and so on. We term the two partitions defined above as the forward LOS partition (FLP) and backward LOS partition (BLP), respectively. The idea of L94 is useful when modelling ecological data since, as mentioned above, sampling is typically performed only at discrete times, $t_1,\dots,t_K$. An application of this partition is presented in Section \ref{sec:cjs}.

In some cases, it is useful to build the partition in a single step by splitting using all the sampling times simultaneously. That is, we split $\Omega$ in $(B_0,\dots,B_K)$, where $B_i = (t_i,t_{i+1})$ and $t_0 = -\infty$, $t_{K+1} = + \infty$, and the masses are assigned according to a Dirichlet distribution. This choice implies negative correlation between probabilities of any two disjoint sets. This partition is termed the uniform partition (UP).

\subsection{Example: Cormack-Jolly-Seber}
\label{sec:cjs}

In this section we demonstrate how the CJS model, briefly described in Section \ref{sec:sampling}, can be expressed in a PT framework. Following the original CJS model formulation, we define the model for the data conditioning on the time of first capture.  

\begin{figure}[h]
\hspace{-0.2in}
  \includegraphics[width=16cm]{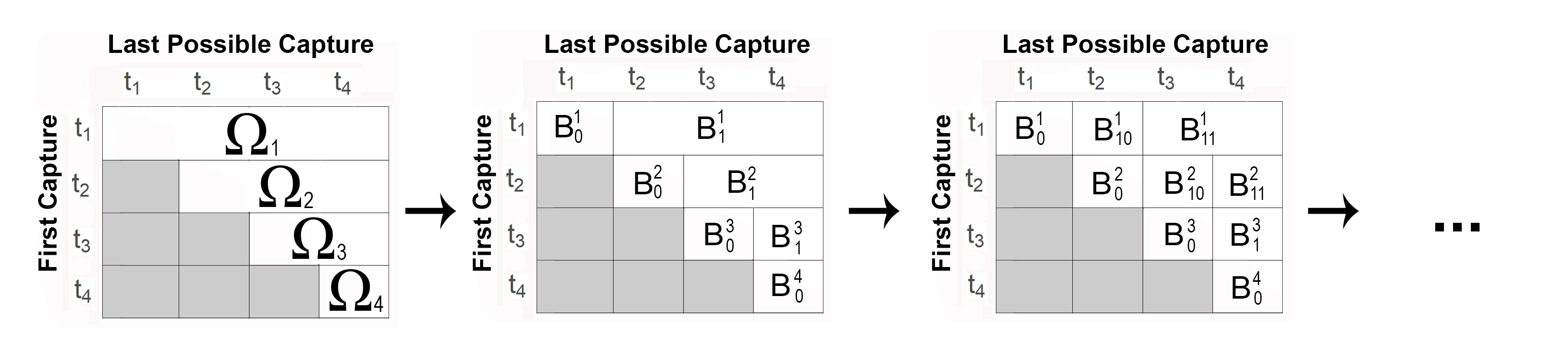}
  \caption{Partitions of the PT in the CJS case. The first level corresponds to the sample spaces $\Omega_k = (t_k,\infty)$, consisting of the individuals with first capture at time $t_k$. In the second level, each sample space  is split into $B^k_0 = (t_k,t_{k+1})$ and $B^k_1 = (t_{k+1}, \infty)$ and so on. The sets $B^k_{\mathbf{1}_j 0}$ and $B^k_{\mathbf{1}_j 1}$ correspond to the individuals first caught at time $t_k$ that are present at the site at time $t_{k + j}$ and exit and remain, respectively, at the site until the next sampling time, $t_{k+j+1}$.}
\label{fig:cjs_part}
\end{figure}

We first arrange all the marked individuals in a set of vectors $  \textbf{n}^k = (n^k_1, \dots, n^k_{K - k + 1})$, where $n^k_{j}$ indicates the number of individuals with first capture at time $t_k$ and remaining at the site for $j$ sampling occasions or, equivalently, with last \textit{possible} capture at time $t_{k+j-1}$. Next, we assume a distribution $G_k$ on the time of last possible capture for the individuals belonging to vector $\textbf{n}^{k}$ or, equivalently, for the individuals with first capture at time $t_k$. The sample space for $G_k$ is $\Omega_k = (t_k, \infty)$, as exit time is left truncated by the time of first capture. We assume a PT prior for each $G_k$, with partition taken to be the FLP. According to the FLP, the sample space $\Omega_k$ is split into $B^k_{0} = (t_k,t_{k+1})$ and $B^k_{1} = (t_{k+1}, \infty)$, next $B^k_{1}$ is further split into $B^k_{10} = (t_{k+1},t_{k+2})$ and $B^k_{11} = (t_{k+2}, \infty)$ and so on.  The last level of the partition corresponds to the grid defined by the capture times, $(t_k,t_{k+1},\dots,t_K)$, and thus it uniquely determines the probabilities of each cell. A visual representation of the sets of the sample space is shown in Fig. \ref{fig:cjs_part}.  The $K$ partitions have different depths as for the individuals with first capture at time $t_k$, the partition is built only for $K - k + 1$ levels. 


For illustration purposes, we assume that individuals are of age $1$ when first caught at time $t_k$ but will later show how to relax this assumption. We define $V^k_{j}$ to be the probability of an individual aged $1$ when first caught at time $t_k$ exiting before time $t_{k + j}$ conditional on being present and of age $j$ at the site at time $t_{k + j - 1}$. The $V^k_j$ can also be seen as the ratio of masses $\frac{G_k(B_{ \mathbf{1}_j 0})}{G_k(B_{\mathbf{1}_j})}$ (where $\mathbf{1}_j = (\underbrace{1,\dots,1}_{j \text{ times}})$) in the PT.

As mentioned in Section \ref{sec:sampling}, in the standard CJS model, $\phi_{j,a}$ corresponds to the probability that an individual present and aged $a$ at time $t_j$ remains until time $t_{j+1}$.  As individuals first caught at time $t_k$ are of age $j$ at time $t_{k+j-1}$, according to our model $V^{k}_j = 1 - \phi_{k+j-1,j}$.

The advantage of using a PT prior is that different assumptions regarding the model for parameter $\phi_{j,a}$ can be considered by sharing the r.v.s $V^k_{j}$ across the different PTs. We list below a number of assumptions that can be made and demonstrate the concept by writing the probability of capture history $0101$, $\mathbb{P}(0101)$, according to each assumption. 

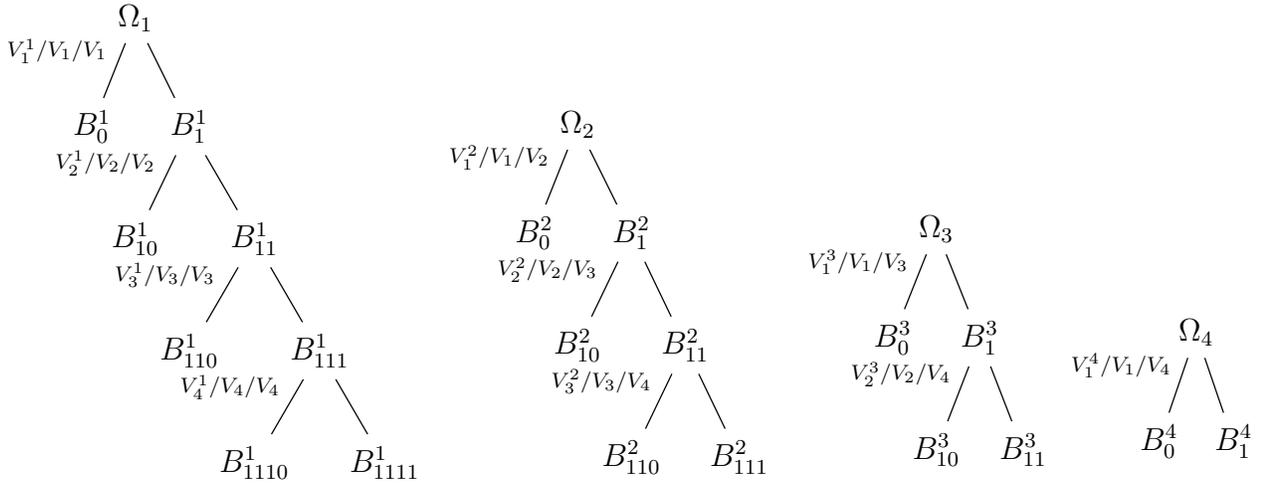
\begin{figure}[h]
\hspace{-1.1cm}
\begin{minipage}{0.34\textwidth}
$$
\begin{tikzpicture}[baseline= (a).base]
\node[scale=1] (a) at (0,0){
\begin{tikzcd}[column sep = -10pt]
& \Omega_1 \arrow[ld, no head, swap, "V^1_{1} / V_{1} / V_1"] \arrow[rd, no head]    & & & &  \\
B^1_{0}  & & B^1_{1} \arrow[ld, no head, swap, "V^1_{2} / V_{2} / V_2"] \arrow[rd, no head] & & & & &  \\
&  B^1_{10} & &  B^1_{11} \arrow[ld, no head, swap, "V^1_{3} / V_{3} / V_3"] \arrow[rd, no head]  &  &  \\
& &     B^1_{110} & &  B^1_{111} \arrow[ld, no head, swap, "V^1_{4} / V_{4} / V_4"] \arrow[rd, no head] & \\
& & &   B^1_{1110} & & B^1_{1111}
\end{tikzcd}
};
\end{tikzpicture}
$$
\end{minipage}
\begin{minipage}{0.28\textwidth}
\vspace{.6cm}
$$
\begin{tikzpicture}[baseline= (a).base]
\node[scale=1] (a) at (0,0){
\begin{tikzcd}[column sep = -10pt]
& & & & &  \\
 & & \Omega_2 \arrow[ld, no head, swap, "V^2_{1} / V_{1} / V_2"] \arrow[rd, no head] & & & & &  \\
&  B^2_{0} & &  B^2_{1} \arrow[ld, no head, swap, "V^2_{2} / V_{2} / V_3"] \arrow[rd, no head]  &  &  \\
& &     B^2_{10} & &  B^2_{11} \arrow[ld, no head, swap, "V^2_{3} / V_{3} / V_4"] \arrow[rd, no head] & \\
& & &   B^2_{110} & & B^2_{111}
\end{tikzcd}
};
\end{tikzpicture}
$$
\end{minipage}
\begin{minipage}{0.22\textwidth}
\vspace{1.15cm}
$$
\begin{tikzpicture}[baseline= (a).base]
\node[scale=1] (a) at (0,0){
\begin{tikzcd}[column sep = -10pt]
& & & & &  \\
 & &  & & & & &  \\
&  & &  \Omega_{3} \arrow[ld, no head, swap, "V^3_{1} / V_{1} / V_3"] \arrow[rd, no head,]  &  &  \\
& &     B^3_{0} & &  B^3_{1} \arrow[ld, no head, swap, "V^3_{2} / V_{2} / V_4"] \arrow[rd, no head, ] & \\
& & &   B^3_{10} & & B^3_{11}
\end{tikzcd}
};
\end{tikzpicture}
$$
\end{minipage}
\begin{minipage}{0.15\textwidth}
\vspace{3.9cm}
$$
\begin{tikzpicture}[baseline= (a).base]
\node[scale=1] (a) at (0,0){
\begin{tikzcd}[column sep = -10pt]
&   \Omega_{4} \arrow[ld, no head, swap, "V^4_{1} / V_{1} / V_4"] \arrow[rd, no head,]  &  \\
B^4_{0} & &  B^4_{1} 
\end{tikzcd}
};
\end{tikzpicture}
$$
\end{minipage}
\caption{Structure of the trees for the CJS model. The r.v. assigning the ratio of the masses of two sets for each branch is represented on the left branch. The r.v.s are in the order \textit{unconstrained}/\textit{age-dependent}/\textit{time-dependent} case. The mass assigned to the set in each right branch is always one minus the mass assigned to the set in the corresponding left branch.}
\label{fig:cjstrees}
\end{figure}

\begin{itemize}
\item \textit{Constant case}: $\phi_{j,a} \equiv \phi \Rightarrow V^k_{j} \equiv V \ \forall k,j$. $\mathbb{P}(0101) = \phi^2 p(1-p)$.
\item \textit{Age-dependent case}: $\phi_{j,a} \equiv \phi_a \Rightarrow V^k_{j} \equiv V_j \ \forall k$, that is equivalent to assuming that $G_k \equiv G \ \forall k$.  $\mathbb{P}(0101) = \phi_{1} \phi_2 p(1-p)$. 
\item \textit{Time-dependent case}: $\phi_{j,a} \equiv \phi_j \Rightarrow V^k_{j} \equiv V_{k+j}$. This is equivalent to assuming that the distribution of $G_k(x)$ is the same as the distribution of $G_{k-1}(x | x \in B_{k-1,1}) \ \forall k$. $\mathbb{P}(0101) = \phi_{2} \phi_3 p(1-p)$. 
\item \textit{Unconstrained case}: The most general case is obtained by assuming that all the $V^k_{j}$ are different. Although this model is never fitted in practice because it is overparametrised, we can still write $\mathbb{P}(0101)$ as $\phi_{2,1} \phi_{3,2} p(1-p)$.
\end{itemize}
These assumptions are depicted in Fig. \ref{fig:cjstrees}. The likelihood of the data and the MCMC algorithm used to sample from the posterior distribution are respectively in Section $3.1$ and $6.1$ of the supplementary material.

\section{Replicate PT}
\label{sec:rpt}

In the CJS model defined in the previous section, different models are obtained by assuming that parts of different trees are ``replicated'' across different branches. The same idea will also be used in the next sections, and hence we provide a formal definition of the RPT. The definition we provide is for a single tree but it is easy to extend to the case of multiple trees.

Let $\Sigma$ be a rule to generate a sequence of partitions from an initial seed set. For example, in the CJS example the rule $\Sigma$ corresponds to FLP and the seed sets are $\Omega_1,\dots,\Omega_K$. If $G$ is a distribution having a PT prior with partition $\Pi$, we say that $G$ has a Replicate PT (RPT) structure across two sets $B_{\epsilon}$ and $B_{\epsilon'}$ of the partition $\Pi$ if the following constraints hold:
\begin{itemize}
\item The partitions of the trees starting from $B_{\epsilon}$ and $B_{\epsilon'}$ are generated according to the same rule $\Sigma$, although the two partitions could be stopped after a different number of steps;
\item For all $\epsilon_1, \dots, \epsilon_m$ , $V_{\epsilon, \epsilon_1, \dots, \epsilon_m} = V_{\epsilon', \epsilon_1, \dots, \epsilon_m}$.
\end{itemize}
The first condition states that the partitions of the trees starting from $B_{\epsilon}$ and $B_{\epsilon'}$ are the same for a number of steps, while the second states that the r.v.s in the two trees are the same. The assumption that the r.v.s are shared across different parts of the tree does not change the conjugate scheme described in Section \ref{PTsection}. The definition is easily extended to the case where more than two parts of a tree are shared.


Some of the assumptions of the CJS model, depicted in Fig. \ref{fig:cjstrees}, can be expressed in terms of the RPT just defined.  The age-dependent case is obtained assuming that the trees have a replicate structure across the sets $\Omega_1, \dots, \Omega_K$, while the time-dependent case is obtained assuming a replicate structure across the pairs of trees with seed sets: $\Omega_2$ and $B^1_{1}$, $\Omega_3$ and $B^2_{1}$, $\Omega_4$ and $B^3_{1}$ and so on.

For ease of exposition, in all the case studies shown later, we only consider the most commonly employed constraint in each case. However, similarly to the CJS case described above, different RPT structures can be employed in order to assume different constraints if required. 


\section{Bivariate PT models}

\subsection{Bivariate partitions}
\label{sec:bivp}

In this section, we extend the model for CR data conditional on the time of first capture of Section \ref{sec:cjs} to two scenarios: CR without conditioning on first capture and RR. To highlight the connection with D18, we briefly describe their model here. D18 define a model for count data by introducing a matrix $\{ n_{fl} \}$ of the individuals with first possible detection at time $t_f$ and last possible detection at time $t_l$. Therefore, inferring matrix $n_{fl}$, which also includes information on the latent entry intervals, gives rise to an estimate of the entry and exit pattern and the size of the population.

In the CR and RR extensions, we do not condition on first capture as in the CJS but we model the entry intervals of the individuals, and thus instead of working with a set of vectors $\{ n^k_{j} \}$ we will introduce a set of matrices $\{ n^k_{fl} \}$, similarly to D18. We note that, as opposed to D18, in the case of uniquely identifiable individuals, as in CR and RR data, the time of first capture $t_k$ is also needed to obtain an analytical expression for the likelihood function. However, the downside of adding another dimension is that it increases the number of parameters of the model. Hence, we share information between individuals with different times of first capture $t_k$ by using the RPT structure, in a similar fashion as this was performed for the CJS model.

The RR and CR models are built by using the same set of matrices $\{ n^k_{fl} \}$ to summarise the entry and exit patterns, but we demonstrate how different prior distributions can be elicited depending on whether prior knowledge is available on entry and exit (CR) or LOS (RR).


In order to define the aforementioned models, we need to define partitions for distributions on $\mathbb{R}^2$, as we consider joint distributions on entry and exit or on entry and LOS. Partitions for distributions on $\mathbb{R}^2$ can be defined by using the schemes defined in Section \ref{sec:unip} separately for each dimension.  We note that do not consider partitions built by considering splits in more than one dimension at a time, such as in \citet{paddock2003randomized}, for two reasons. First, building univariate partitions sequentially for each dimension makes it easier to center the PT on standard parametric models. For example, if the distribution of interest is the bivariate entry and exit distribution, building a sequential model by splitting first according to entry and then according to exit assumes a centering model where exit is independent on entry. Secondly, interpretation from the model by considering joint splits becomes more difficult. In fact, inference on the conditional distribution of one dimension given the other is not straightforward to obtain anymore in this case.

A useful bivariate partition can be constructed by applying first FLP in one dimension and next BLP in the other. We denote this partition as entry and exit bivariate partition (EEBP), and the scheme is depicted in Fig. \ref{fig:adsp}. A formal definition of the partition is presented in Section $2.1.1$ of the supplementary material.



This partition is useful when the entry and exit distribution is of interest, as the marginal distribution of entry is expected to have more variance in the left tail, while for the marginal exit distribution more variance is assumed in the right tail. This partition is used in D18, who built a model for CD based on a PT. The model is further extended to CR data in Section \ref{sec:joint}.

When no prior information is available on the variance of the entry or exit distributions, the forward/backward splits can be replaced by the uniform split (UP), described in Section \ref{sec:unip}, shown in Fig. $3$ of the supplementary material and used in Section \ref{sec:moths}.


%

The two previous partitions allow us to elicit prior information on the entry and exit distribution. However, as mentioned above, in some applications, prior information is available on the LOS distribution instead. For eliciting a prior on the LOS distribution, we define a bivariate partition based on the idea of L94 but split the sample space w.r.t. the LOS and then w.r.t the entry intervals, as shown in Fig. \ref{fig:3}. The definition is given in Section $2.1.2$ of the supplementary material.  This partition is termed bivariate LOS partition (BivLP) and an application is presented in Section \ref{sec:ring}.

\subsection{Joint analysis of capture-recapture and count-data}
\label{sec:joint}

Recent work \citep{barker2018reliability} casted doubt on whether CD alone can be used to provide reliable estimates of population size and suggested that analysis of CD should be augmented by other types of data, such as CR. Therefore, in this section, we demonstrate how to extend D18 to perform a joint analysis of CR and CD using an RPT. As introduced in the previous section, we define the set of matrices $\{ \textbf{n}^k \}= \{ n^k_{fl} \}_{f=1,\dots,K+1, l=1,\dots,K+1}$, where $n^k_{fl}$ denotes the number of individuals with first possible capture at time $t_{f}$, first capture at time $t_k$ and last possible capture at time $t_l$ (clearly, $n^k_{fl} = 0$ for $f > k$). For $k = 0$, we define $ \textbf{n}^0$ as the matrix containing the individuals never captured, but possibly detected and hence contributing to the CD. 

For each matrix $ \textbf{n}^k $, we define a distribution $G_k$ on the probabilities $\omega^k_{fl}$ of an individual belonging to each cell of the matrix $\{ n^k_{fl} \}$, and we assume a PT prior for each $G_k$ using the EEBP partition of Section \ref{sec:bivp}. We make the assumption that the distribution of entry and exit intervals is the same for each matrix $ \textbf{n}^k $. If $\Omega_k$ is the sample space for the individuals with first capture at time $t_k$, this is achieved by assuming a RPT prior across the sets $\Omega_1,\dots,\Omega_K$, as in Section \ref{sec:cjs}. Moreover, we also assume that the exit distribution is independent of the entry interval, which is achieved by assuming, for each slice $k$, a RPT structure across the sets $B_0 = \{ (x, y) : x < t_{1} \}$, $B_1 = \{ (x, y) : t_1 < x < t_{2} \}$, $\dots$, $B_{k-1} = \{ (x, y) : t_{k-1} < x < t_{k} \}$ (see level $K$ of Fig. \ref{fig:adsp} for a visual representation of the sets). 

We note that each matrix is defined on a different sample space, as $\textbf{n}^k$ is defined only for $f \le k, l > k$, because individuals in the $k$-th slice have to enter before the $k$-th sampling occasion and exit after. A representation of one slice is shown in Fig. $4$ of the supplementary material.

We define the sum of the elements of matrix $\textbf{n}^k$ as $D_k$. However, while $D_k$ is known for $k=1,\dots,K$ (as it is equal to the number of individuals with first capture at time $t_k$), $D_0$ is unknown as it corresponds to the number of individuals never captured. Given the number of individuals $D_k$, the number of individuals in each cell of the matrix is distributed as a multinomial with sample size $D_k$ and vector of probabilities $\omega^k_{fl}$. The total number of individuals $N$ is then equal to $D_0 + D$, where $D = \sum_{k=1}^K D_k$ is the number of caught individuals. We also define as $N_j$ the number of individuals available for capture/detection at time $t_j$, conditional on the latent matrices $n^k_{fl}$. Assuming a $\text{Poisson}(\omega)$ prior distribution for $N$, the model can be summarised as 
$$
\begin{cases}
C_j \sim \text{Binomial}(N_j, p_D)  \hspace{2cm}
N_j = \sum_{k=0}^j \sum_{l=j}^K \sum_{f=1}^{l-1} n^k_{fl}  \\
n^k_{fl} \sim \text{Multinomial}(D_k, \omega^k_{fl}) \hspace{2cm}
N \sim \text{Poisson}(\omega) 
\end{cases}
$$
where $p_D$ and $p_C$ are the probability of detecting and capturing an individual, respectively. The probability of the capture histories of the individuals with first capture at time $t_k$, conditional on the matrix $\textbf{n}^k$, is presented in Section $3.2$ of the supplementary material. 



\subsubsection{Case study}
\label{sec:cs_op}

We apply our model to CD and CR data collected on a population of Italian spoonbills (\textit{Platalea leucorodia}) collected in the southern Po delta, in North-East Italy. Birds are captured as chicks in previous years and are uniquely marked. The CR dataset is collected by resighting adult birds through photographs obtained using camera traps and by visiting their nests on eight separate sampling occasions. No attempt is made to mark new adult individuals and instead only previously marked individuals can be resighted. In addition, unmarked birds are detected on each sampling occasion. In this case, there are fewer than $40$ resighted individuals, so we choose to make inference by modelling individual entry and exit intervals for all the marked individuals resighted. The full details of the model are given in Section $5.1$ the supplementary material. 

The posterior means of the entry and exit cumulative distribution functions (CDFs) are shown in Fig. $6$ (a) of the supplementary material. The CDF of the departure does not reach $1$ within the study period since some birds remain at the site using the colony as a roost. Similarly, the CDF of the arrival intervals is at around 0.4 at the start of the study, suggesting that a large proportion of birds are present at the site when sampling starts, with over 80\% of birds estimated to have arrived by the second sampling occasion toward the end of April. The posterior distributions of the two population sizes and resighting probabilities are shown in Fig. $6$ (b) of the supplementary material. The camera trap resight probability, $p_C$, is slightly lower than the resight probability through nest visits, $p_R$, since cameras are pointed toward the nest, where it is unlikely to see floaters and prospectors. The difference between the population sizes of marked and unmarked birds is due to the fact that not all the chicks are marked each year, and out of those marked, only a small proportion return to breed as adults. In fact, local recruitment rate is thought to be around $0.12$, while the proportion of immigrants on total number of recruits is ranges from $0.49$ to $0.83$ \citep{tenan2017conspecific}.




\subsection{Ring-Recovery}
\label{sec:ring}

In this section, we show how to build a model for RR data using a RPT prior.  Similarly to what was mentioned in Section \ref{sec:bivp}, in order to jointly model entry and exit patterns we work with a set of matrices $\textbf{n}^k = \{ n^k_{u_f u_l} \}$, where in this case $n^k_{u_f u_l}$ represent the number of individuals marked on sampling occasion $k$ that were present in the population for $u_f$ sampling occasions before being marked and remained for $u_l$ sampling occasions after being marked. We assume that individuals marked at time $t_k$ could have entered the population for up to $U$ sampling occasions before they were marked and could have stayed for up to $U$ sampling occasions after being marked, and thus we assume the dimension of each matrix $\textbf{n}^k$ is $(U+1) \times (U+1)$.

For each matrix $\textbf{n}^k$, we assume a PT prior for the probabilities $\omega^k_{u_f u_l}$ for an individual to belong to each cell $n^k_{u_f u_l}$, with partition taken to be the BivLP partition built in Section \ref{sec:bivp}. A visual representation of the partition for a matrix is presented in Fig. $5$ of the supplementary material. 

Similarly to Section \ref{sec:cjs}, we make the assumption that the survival probability is age-dependent, that is, the probability of an individual surviving until the next sampling occasion depends only on the age of the individual and not on the sampling occasion. If we define $\Omega_k$ to be the sample space for the individuals marked at time $t_k$, this assumption can be achieved by assuming an RPT across $\Omega_1,\dots,\Omega_K$. This forces the probabilities $\omega^k_{u_f u_l}$ to be the same for each $k$, that is $\omega^k_{u_f u_l} \equiv \omega_{u_f u_l}$. If we define $\phi_{j,a}$ as the probability of an individual of age $a$ that is present at time $t_j$ to remain until the following sampling occasion, assuming an RPT prior over the different slices is equivalent to assuming that $\phi_{j,a} \equiv \phi_a$.


The number of individuals marked at time $t_k$ that can be recovered at time $t_{k + j}$ can be obtained by the matrix $n^k_{u_f u_l}$ as $\sum_{u_f=0}^U n^k_{u_f j}$ by summing over the different entry times. Hence, the number of individuals marked in year $t_k$ recovered dead in year $t_{k + j}$, $R_{k,k+j}$, is distributed as a $\text{Binomial}(\sum_{u_f=0}^U n^k_{u_f j}, \lambda)$. Assuming also a Beta prior for the recovery probability $\lambda$, the model can be summarised as
\begin{equation}
\label{eq:rr1}
\begin{cases}
R_{k,k+j} \sim \text{Binomial}(\sum_{u_f=0}^U n^k_{u_f j}, \lambda)  \hspace{2cm} n^k_{u_f u_l} \sim \text{Multinomial}(m_k, \omega_{u_f u_l}) \\
\{ \omega_{u_f u_l} \} \sim \text{PT}( \Pi, \{ \alpha \})  \hspace{4cm}
\lambda \sim \text{Beta}(a_0, b_0) 
\end{cases}.
\end{equation}

\subsubsection{Case study}

We apply our model to a dataset collected in Minnesota, USA. A total of $100127$ female mallards were banded throughout the course of the study, which lasted 51 years. Marking occurs from July to September, while recoveries occur during the hunting season immediately following marking, from September to January. Newly caught individuals can be recognized as juveniles if their age is less than $1$ year at the time of capture and as adults otherwise. Therefore, individuals caught as juveniles that have been recovered are of known age at year of death. The entry and exit of the individuals correspond in this case to births and deaths, while the sampling occasions correspond to the markings and recoveries. The full details of the model are given in Section $5.2$ of the supplementary material.

The posterior mean of the survival probabilities is presented in Fig. $7$ of the supplementary material. The result agrees with the general pattern observed for bird populations, with survival being lower in very young and older ages. A similar pattern was observed by \cite{mccrea2013age} when analysing RR data for mallards (\textit{Anas platyrhynchos}) and by \cite{jimenez2019estimating} for blackbirds (\textit{Turdus merula}), but we note here that we have not employed a parametric curve to enforce this pattern, and instead if has been completely driven by the data. As expected, uncertainty increases for older ages because of the sparseness of the data. The posterior mean of the recovery probability is equal to $0.123$, (95\% PCI: $0.121-0.125$) which is in line with findings of similar studies \citep{mccrea2012conditional}.

\section{The hierarchical Logistic PT}
\label{sec:hpt}

The RPT defined in Section \ref{sec:rpt} allows us to impose constraints across different distributions or conditional distributions by assuming the same r.v.s across different trees or across different branches of the same tree, which in the case of density estimation forces the different distributions to be the same. This assumption is however too restrictive in many real applications, as for example when jointly modelling data collected across different years or sites. In these cases, it may be more reasonable to employ a hierarchical approach, which enables us to borrow strength across the different datasets but without assuming that the distributions describing the different datasets are exactly the same. 

In the context of data collected at different units, \citet{christensen2017bayesian} propose a Hierarchical PT by assuming a PT prior in each unit, where each PT is centered on a common baseline PT prior. However, because of the lack of conjugacy of a Beta prior with a Beta likelihood, the sampling scheme relies on quadrature techniques. To employ a hierarchical approach while still performing inference with a Gibbs sampler, we suggest an alternative approach based on the logistic PT (LPT) of \citet{jara2011class}.   The LPT is defined by replacing the Dirichlet distributions for the $V_{\epsilon_1 \dots \epsilon_m}$ in the definition of the PT with multinomial logistic normal distributions.


If $G_1,\dots,G_s$ are the distribution in each unit and $V^s_{\epsilon_1 \dots \epsilon_m}$ are the conditional distributions assigning the mass in each split in the $s$-th unit, that is $V^s_{\epsilon_1 \dots \epsilon_{m}, i} = \mathbb{P}(x \in B_{\epsilon_1 \dots \epsilon_{m}, i} | x \in B_{\epsilon_1 \dots \epsilon_{m}})$, we define the hierarchical model as
$$
\begin{cases}
V^s_{\epsilon_1 \dots \epsilon_{m}, i} =
\begin{cases}
\frac{ \exp{ \left( \beta^s_{\epsilon_1 \dots \epsilon_{m}, i} \right) }}{ 1 + \sum_{k = 0}^{K-2} \exp{ \left( \beta^s_{\epsilon_1 \dots \epsilon_{m}, k} \right) } } \hspace{3cm}  i = 0,\dots,K-2 \\
\frac{ 1 }{ 1 + \sum_{k = 0}^{K-2} \exp{ \left( \beta^s_{\epsilon_1 \dots \epsilon_{m}, k} \right) } } \hspace{3cm}  i = K-1
\end{cases} \\
\beta^s_{\epsilon_1 \dots \epsilon_{m}} \sim \text{N}(\mu^0_{\epsilon_1 \dots \epsilon_{m}}, \sigma_{\epsilon_1 \dots \epsilon_{m}}^2) \hspace{4cm} s = 1,\dots,S \\
\mu^0_{\epsilon_1 \dots \epsilon_{m}} = 
\text{logit} \left( \frac{G_0(B_{\epsilon_1 \dots \epsilon_m} | \eta)}{G_0(B_{\epsilon_1 \dots \epsilon_{m-1}} | \eta)} \right)
 \hspace{4cm}
\eta \sim P_0
\end{cases}
$$
where $\mu^0$ are the parameters of a PT centered on a distribution $G_0(\cdot | \eta)$ and an additional hyperprior $P_0$ is assumed on $\eta$. A set of PTs following this structure will be termed Hierarchical Logistic PT (HLPT). If $\sigma_{\epsilon}^2 \rightarrow 0 \ \forall \epsilon$, we have $\beta^s_{\epsilon} \equiv \beta_{\epsilon}$ and hence the distributions for each group are the same, reducing to the case of the RPT.


The LPT is defined in terms of a logistic transform of normal distributions and hence a Gibbs sampler, based on the Polya-Gamma scheme \citep{polson2013bayesian}, is available, details of which can be found in Section $6.4$ of the supplementary material. 

\subsection{A hierarchical model across different datasets}
\label{sec:hierarchical}

In this section we extend D18 by defining a model for related datasets using the HLPT. The data consist of $S$ different datasets which, for ease of exposition, we take to be CD, but the same rationale applies for any other sampling schemes described before. 

We define $C_{js}$ to be the number of individuals detected at time $t_j$ in the $s$-th dataset. We arrange all the individuals in $S$ matrices $ \textbf{n}^s = \{ n^s_{fl} \}_{f=0,\dots,K, l=0,\dots,K}$, where $n^s_{fl}$ is the number of individuals in dataset $s$ having first and last possible detection at times $t_f$ and $t_l$, respectively. For the $s$-th dataset, individual entry and exit times are drawn from a Poisson process with intensity $\nu_s = \omega_s \times G_s$, where $\omega_s$ is the overall intensity and $G_s$ is the normalized density.  In order to borrow information between different datasets, the normalized densities $(G_1,\dots,G_S)$ are assumed to follow a HLPT prior, where the partition is taken to be the EEBP of Section \ref{sec:bivp}. 

Assuming also that each individual in dataset $s$ can be detected with probability $p_s$, the model can be summarised in the following hierarchical structure
$$
\begin{cases}
C_{js} \sim \text{Binomial}(N_{js}, p_s)  
\hspace{3cm}
N_{js} = \sum_{l=j}^K \sum_{f=0}^{j-1} n^s_{fl}  \\
 n^s_{fl} \sim \text{Poisson}(\omega_s \times \omega^s_{fl}) 
 \hspace{3cm}
\{ \omega^s_{fl} \} \sim \text{HLPT}( \eta, \sigma) \\
\eta \sim \text{P}_0 
\hspace{6cm}
\end{cases}
$$
where $N_{js}$ is the number of individuals available for sampling on occasion $j$ and dataset $s$.

\subsubsection{Case study}
\label{case:newts}

We consider a dataset of weekly detections of $3$ species of newts separated by sex, resulting in $6$ different sets of CD, collected at a local site at the University of Kent in $2014$. Newts are usually sampled weekly during their breeding period using aquatic traps. However, some visits might be missed on occasion because of weather conditions. All three species migrate to ponds in late winter-early spring for breeding, and leave the breeding site in late spring or summer to spend the rest of the year on land. The data are shown in Figure $8$ of the supplementary material.




The posterior estimates of the entry and exit distributions for the $6$ datasets, shown in Fig. $9$ of the supplementary material, suggest that great crested newts tend to arrive and depart later than smooth or palmate newts. Additionally, males arrive earlier, while females depart later, since they need to complete egg-laying. All species are found to have similar detection probabilities with overlapping credible intervals, with the exception of palmate newts that have a slightly lower detection probability, as shown in Fig. $10$ of the supplementary material.

\subsection{Modelling long time-series of ecological data}
\label{sec:moths}

The last scenario we consider is a long time series of CD collected daily. The challenge that arises in this case is that the sampling period is considerably longer than the period during which individuals of the species are present at the site. Therefore, building a partition by considering all $K$ intervals would increase the computational complexity without necessarily providing a better description of the data. A naive modelling approach is to merge sampling occasions in order to obtain a shorter time series. However, the decision of how to merge is arbitrary and the same merging may not be meaningful for the duration of the study period. Instead, we let the model infer which sampling occasions to merge using the Optional Polya Tree of \citet{wong2010optional}. 

As in the model of Section \ref{sec:hierarchical}, we build a prior distribution on the number of individuals present on each sampling occasion by assuming a PT prior on the individual entry and exit intervals. The partition of the PT is built using the UP of Section \ref{sec:bivp}, where the splits are performed according to some nested periods of time, which for example can be taken as the natural scales at which data are collected, such as months, weeks and days. Using the OPT allows us to stop partitioning further, and hence the choice of the UP will lead the partition to end in one of these periods of time.  

We divide the overall study period in $K_0$ periods of time of equal length, $\Omega_1,\dots,\Omega_{K_0}$, called $0$-periods, and we assume that individuals enter and exit only in the same $0$-period. Next, for each $0$-period, $\Omega_{i_0}$, we define a distribution $G_{i_0}$ on the entry and exit intervals of the individuals. The $0$-periods can be thought of as the periods of time in which the entry and exit patterns seasonally repeat, such as the years. Next, as mentioned above, we define a set of splitting rules according to which the sample space of $G_{i_0}$, $\Omega_{i_0}$ is first split into the sets $\Ib^1 = \{ I^1_1,\dots,I^1_{K_1} \}$, and next, each set at level $i$, $I^i_j$, is split into $\Ib^{i+1}_j = (I^{i+1}_{1}, \dots, I^{i+1}_{{K_{i+1}}})$. The sets in $\Ib^i$ are called $i$-periods. Given $\epsilon = (\epsilon_1,\dots,\epsilon_m)$, we define as $T_{\epsilon}$ the set obtained by taking first the $\epsilon_1$-th $1$-period, then the $\epsilon_2$-th $2$-period and so on. For example, the years can be taken as the $0$-periods and the splitting rules can be built using first the months as $1$-periods, weeks as $2$-periods and so on. 

Next, given two sequences of indexes $\epsilon$ and $\delta$, we define as $n^{i_0}(\epsilon, \delta)$ the number of individuals in the $0$-period $\Omega_{i_0}$ entering in $T_{\epsilon}$ and exiting in $T_{\delta}$. We also define as $N^{i_0}_{\epsilon}$ and $C^{i_0}_{\epsilon}$ the number of individuals available for sampling and detected, respectively, in the interval $T_{\epsilon}$ in the $i_0$-th $0$-period. Even if an explicit formula is cumbersome to present, it is straightforward to obtain the latent individuals $N^{i_0}_{\epsilon}$ from the $n^{i_0}(\epsilon, \delta)$. Thanks to this construction, we have defined a prior on the probabilities $\omega^{i_0}(\epsilon, \delta)$ of an individual to belong to cell $n^{i_0}(\epsilon, \delta)$. 
  


\begin{figure}[h]
\hspace{-0.2in}
  \includegraphics[width=17.5cm]{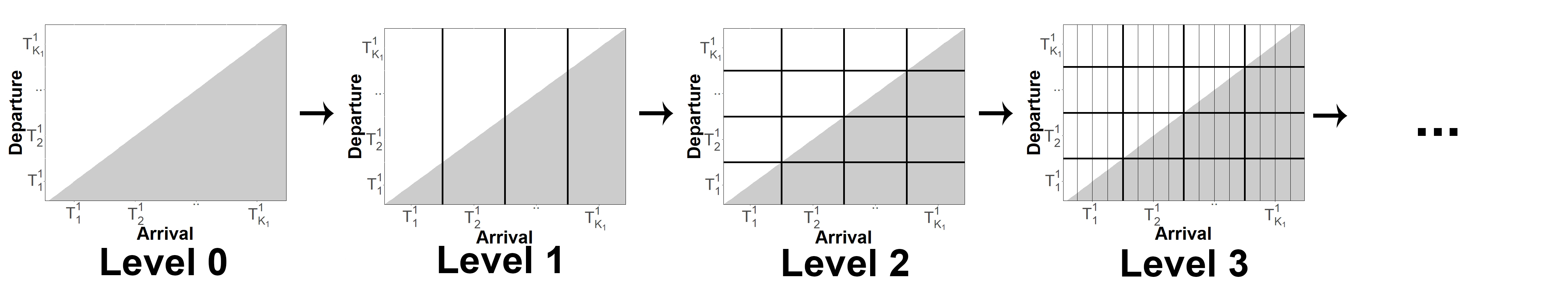}
  \caption{Scheme of partitions for the long time series of CD. The first two splits are performed by splitting first the entry dimension and next the exit dimension according to the $1$-periods, then we split according to the $2$-periods, and so on.}
\label{fig:part_moths}
\end{figure}

Having defined this set of splitting rules, we define the partition of sample space $\Omega_{i_0}$ by first splitting $\Omega_{i_0}$ in $(B^{i_0}_{1},\dots,B^{i_0}_{K_1})$ where $B^{i_0}_{i} = \{ (x,y) \in \Omega_{i_0} , x \in I^1_i \}$, that is, the sample space is split using the dimension of entry according to the $1$-periods. Next, each $B^{i_0}_{i}$ is split into $(B^{i_0}_{i,1},\dots,B^{i_0}_{i,K_1})$, where $B^{i_0}_{i,j} = \{ (x,y) \in B^{i_0}_{i} , y \in I^1_j \}$, that is, each set is now split using the dimension of exit according to again the $1$-periods. The next sets are split in an analogous way. A representation of this scheme is shown in Fig. \ref{fig:part_moths}. For the r.v.s assigning the masses in each level of the partition, we assume an HLPT structure across the years and, similarly to Section \ref{sec:joint}, we use an RPT structure to enforce that individuals arriving in different sets share the same departure distribution. Full details are given in Section $4$ of the supplementary material.

Assuming a Poisson$(\omega)$ prior distribution on the number of individuals $N_{i_0}$ in each $0$-period, the model can be summarised in the following hierarchical structure
$$
\begin{cases}
C^{i_0}_{\epsilon} \sim \text{Binomial}(N^{i_0}_{\epsilon}, p) \\
n^{i_0}(\epsilon, \delta) \sim \text{Poisson}(\omega \cdot \omega^{i_0}(\epsilon, \delta)) \\
\omega^{i_0}(\epsilon, \delta) \sim \text{HLPT}(\mu^0, \sigma, \sigma_0) 
\end{cases}
$$
where $p$ is the probability of detecting an individual.

The downside of this model is that the partition is built for each period, even if no observations are available in some periods. Ideally, we would like to use a modelling approach where we keep partitioning the PT only in the periods of time where the entry and exit density has significant variation. To do so, we borrow ideas from the OPT of \citet{wong2010optional} and allow for different branches of the tree to not partition further. 

Specifically, we assume that in each step, partitioning of the PT into the further level happens with probability $\rho$. We then introduce a variable $S_{B_{\epsilon}}$ which is equal to $1$ if $B_{\epsilon}$ is not partitioned further and $0$ otherwise. The distribution in the set $B_{\epsilon}$ can be expressed recursively from its descendant sets as
$$
G(x | x \in B_{\epsilon}) = S_{B_{\epsilon}} U(x | x \in B_{\epsilon}) + (1 - S_{B_{\epsilon}}) \sum_{i=1}^{K_{\epsilon}} G( x | x \in B_{\epsilon, \epsilon_i}) 1(x \in B_{\epsilon, \epsilon_i})
$$
where $U$ is the uniform distribution and $(B_{\epsilon, \epsilon_1}, \dots, B_{\epsilon, \epsilon_{K_{\epsilon}}})$ are the $K_{\epsilon}$ descendant sets of $B_{\epsilon}$. Hence, when no splitting is required we replace the PT distribution with a uniform distribution.  

\subsubsection{Case study}
\label{case:moths}

We consider a long series of daily CD of two indistinguishable species of moths, Canadian Melanolophia (\textit{Melanolophia canadaria}) and Signate Melanolophia (\textit{Melanolophia signataria}), collected across $5$ years \citep{pickering2015find}. The data are shown in Section $5.4$ of the supplementary material. Detection of the individuals is possible only during the flight period, which is known to happen only in a limited period of time inside the calendar year. 


The posterior means of the entry and exit CDFs are shown in Fig. $12$ of the supplementary material. The exit pattern is practically identical between the different years, with the c.d.f suggesting a smooth exit pattern with a few bigger exit waves around weeks $36$, $40$ and $44$ (start of September). On the other hand, entry patterns present more variation across years. The species are known to have two broods \citep{covell1984field}, which is evident by the two big entry waves each year around weeks 20 and 24 (mid May) for all years apart from 2013, when the first brood emerged much earlier. This is likely due to the unusual warm temperature at the start of $2013$, as Table $1$ in the supplementary material suggests.

\section{Conclusion}
\label{sec:concl}

We introduced a framework for defining models for ecological data on open populations based on the PT prior. The advantage of this framework is that a wide variety of assumptions can be made on the model structure by changing the tree structure, as a result of the flexibility of the PT. We have applied our framework to  different types of commonly collected ecological data, such as CR, CD and RR. We have also introduced the RPT, which allows us to place constraints on the model parameters, and the HLPT, which allows us to impose a hierarchical structure and share information when modelling different datasets. We have also presented a way to infer the partition of a PT using the OPT of \citet{wong2010optional}, which allows us to build trees with depth informed by the data.

It is easy to extend the model to other protocols, as for example removal data \citep{otis1978statistical, matechou2016open}. Removal data are collected by repeatedly visiting the site and removing all caught individuals. In this case, an approach similar to CR and RR can be employed, by assuming that the time of exit is known for the removed individuals and estimating the number of the individuals not removed. 



The HLPT prior was introduced to build hierarchical models across sties. As the model relies on normal structures, it is easy to extend the model while keeping the same posterior inference strategy. For example, it is easy to introduce covariates across the different sites, or to model different sites over time.


Although the OPT allows sampling from the full conditional distribution thanks to the conjugacy property, sampling from the posterior involves recursive calculations across the whole tree. This is computationally inefficient if the tree is large and there are many layers, as this has to be repeated in each iteration. In that case, a better scheme is to prune the tree and in each step of the MCMC only propose to update/add/remove branches of the tree.

In our modelling framework, different ecological assumptions correspond to different trees or changes in the dependence structure of the r.v.s on the tree. The question of evaluating evidence in favour of different assumptions can be addressed using standard Bayesian model selection methods. First, model selection can be performed by evaluating the evidence of each model by using Bayes factors. Alternatively, as a change in the ecological assumption entails a change in the structure of the trees, model selection can be performed by employing an additional prior on these different tree structures and computing the posterior of the structure of trees as part of the parameter space.

\bibliographystyle{biom}
\bibliography{biblio}


\begin{figure}[h]
\hspace{-0.2in}
  \includegraphics[width=16cm]{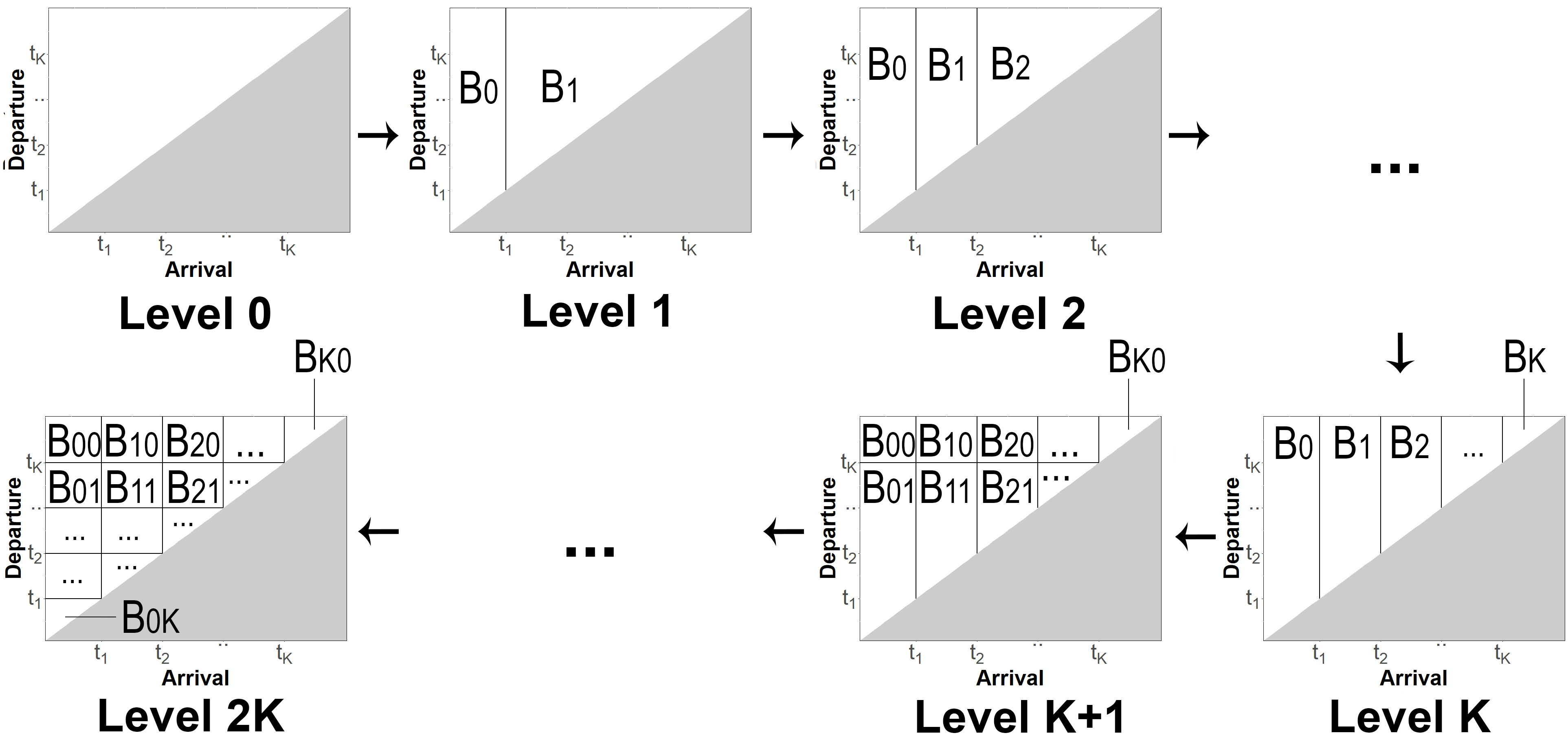}
  \caption{Partition of a PT prior for a distribution defined on $\{ (x,y): (x,y) \in \mathbb{R}^2 , y > x \}$, built first using FLP and next using BLP. For simplicity, we have used the notation $B$ for the sets of the bivariate partition instead of $\tilde{B}$ and omitted the superscripts.}
\label{fig:adsp}
\end{figure}

\begin{figure}[h]
\hspace{-0.2in}
  \includegraphics[width=16cm]{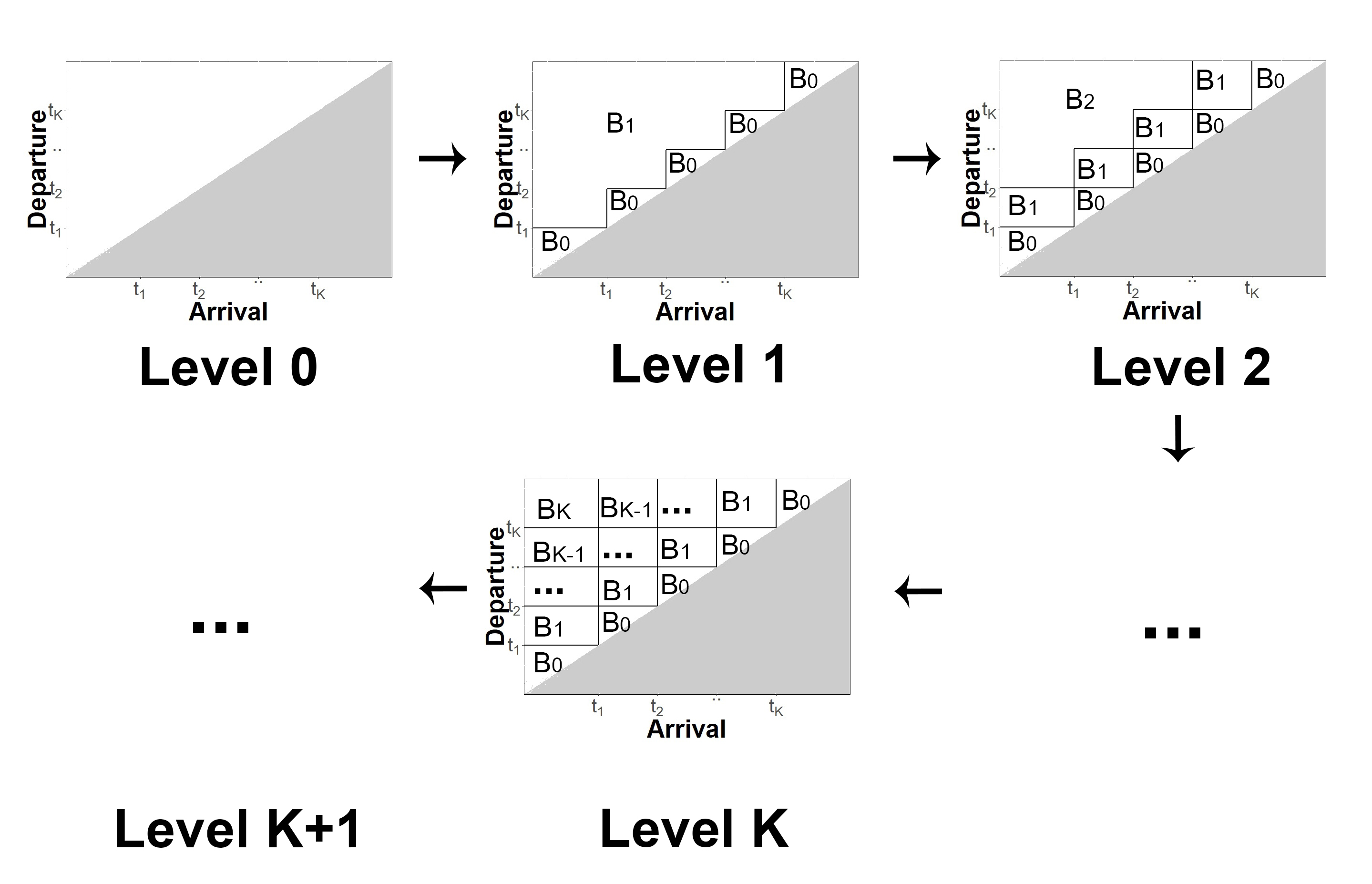}
  \caption{Partition of a PT prior for a distribution defined on $\{ (x,y): (x,y) \in \mathbb{R}^2 , y > x \}$ built using BivLP. The first $K$ levels are built splitting according to the LOS, while the next splits are performed according to the entry dimension.}
\label{fig:3}
\end{figure}

\end{document}


\def\spacingset#1{\renewcommand{\baselinestretch}%
{#1}\small\normalsize} \spacingset{1}


\title{\bf A general modelling framework for open wildlife populations based on the Polya Tree prior - Supplementary material}

  \maketitle

\spacingset{1.5} 

\section{Proof of Proposition 1}

$$
\text{Var}(X_1 X_2) = \text{Var}(X_1) \text{Var}(X_2) + \text{Var}(X_1) \mathbb{E}[X_2]^2 + \text{Var}(X_2) \mathbb{E}[X_1]^2 =
$$
$$
= \frac{\alpha^2}{(2 \alpha)^2 (2 \alpha + 1)} 
\frac{\beta^2}{(2 \beta)^2 (2 \beta + 1)} + \frac{1}{4} \left( \frac{\alpha^2}{(2 \alpha)^2 (2 \alpha + 1)} +
\frac{\beta^2}{(2 \beta)^2 (2 \beta + 1)} \right) =	
$$
$$
= \frac{\alpha^2 \beta^2}{(2 \alpha)^2 (2 \alpha + 1) (2 \beta)^2 (2 \beta + 1)} + \frac{1}{4} \left( \frac{\alpha^2 (2 \beta)^2 (2 \beta + 1) + \beta^2 (2 \alpha)^2 (2 \alpha + 1)}{(2 \alpha)^2 (2 \alpha + 1) (2 \beta)^2 (2 \beta + 1) }  \right) =	
$$
$$
= \frac{\alpha^2 \beta^2 + \alpha^2 \beta^2 (2 \beta + 1) + \alpha^2 \beta^2 (2 \alpha + 1)}{(2 \alpha)^2 (2 \alpha + 1) (2 \beta)^2 (2 \beta + 1)} 
= \frac{\alpha^2 \beta^2 (2 \alpha + 2\beta + 3)}{(2 \alpha)^2 (2 \alpha + 1) (2 \beta)^2 (2 \beta + 1)} =
$$
$$
\text{Var}(X_1) \frac{2 \alpha + 2 \beta + 3}{4(2 \beta + 1)}
$$
and $\frac{2 \alpha + 2 \beta + 3}{4(2 \beta + 1)} \le 1$ as
$$
\frac{2 \alpha + 2 \beta + 3}{4(2 \beta + 1)} \le 1 \iff \beta \ge \frac{\alpha}{3} - \frac{1}{6}
$$

which holds true as $\beta \ge \alpha$.

\newpage

\section{Partitions}

\subsection{Bivariate partitions}

We define here the partitions introduced in Section $5.1$ of the paper.

\subsubsection{EELP partition}

\begin{figure}[H]
\hspace{-0.2in}
  \includegraphics[width=16cm]{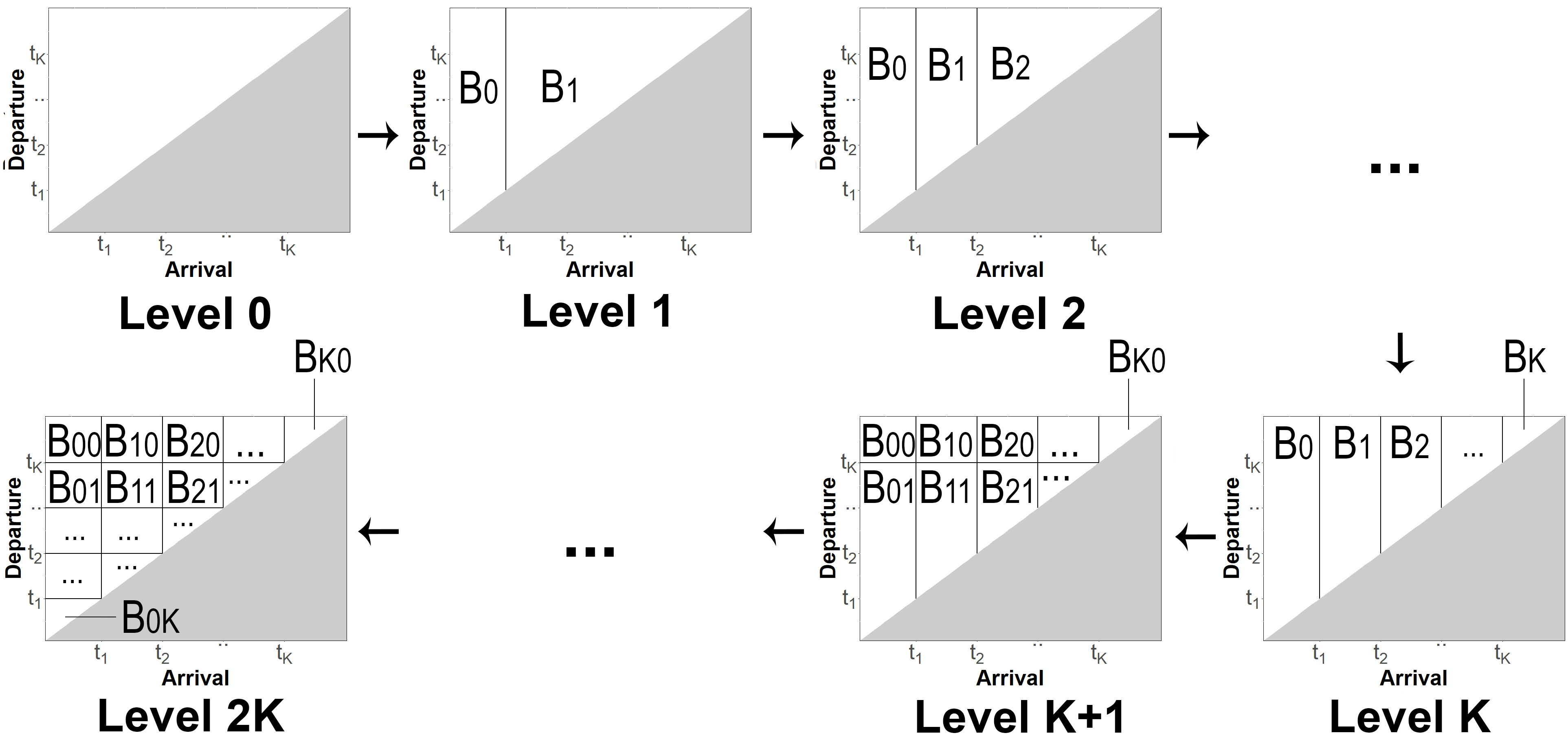}
  \caption{Partition of a PT prior for a distribution defined on $\{ (x,y): (x,y) \in \mathbb{R}^2 , y > x \}$, built first using FLP and next using BLP. For simplicity, we have used the notation $B$ for the sets of the bivariate partition instead of $\tilde{B}$ and omitted the superscripts.}
\label{fig:adsp}
\end{figure}

If $B^1_{\epsilon_1,\dots,\epsilon_k}$ and $B^2_{\epsilon_1,\dots,\epsilon_k}$ are the sets generating with the FLP and BLP partition, respectively, the bivariate partition is defined in the following way:

\begin{itemize}
\item For the first $K$ levels, we define the partition at the $k$-th level as $(\tilde{B}^k_0,\dots,\tilde{B}^k_{k})$ where

$$
\begin{cases}
\tilde{B}^k_{i} = B^1_{ \mathbf{1}_{i-1}, 0} \times (-\infty,\infty)  \hspace{3cm} i = 1,\dots,k-1\\
\tilde{B}^k_{k} = B^1_{\mathbf{1}_{i-1}, 1} \times (-\infty,\infty)  \hspace{3cm} i = k \\
\end{cases}.
$$

This scheme is depicted in the first $K$ levels of Fig. \ref{fig:adsp}. 

\item For the remaining $K$ levels, we define the partition at the $k$-th level as 

 $(\tilde{B}^k_{0,0},\dots,\tilde{B}^k_{0,k},\dots,\tilde{B}^k_{K,0},\dots,\tilde{B}^k_{K_1,k},)$ where

$$
\begin{cases}
\tilde{B}^k_{i,j} = B^1_{ \mathbf{1}_{K}} \times B^2_{ \mathbf{1}_{j-1}, 0}  \hspace{3cm} j = 1,\dots,k-1 \\
\tilde{B}^k_{i,j} = B^1_{ \mathbf{1}_{K}} \times B^2_{ \mathbf{1}_{j-1}, 1}   \hspace{3cm} j = k
\end{cases}.
$$


\end{itemize}

\subsubsection{BivLP partition}

\begin{figure}[H]
\hspace{-0.2in}
  \includegraphics[width=16cm]{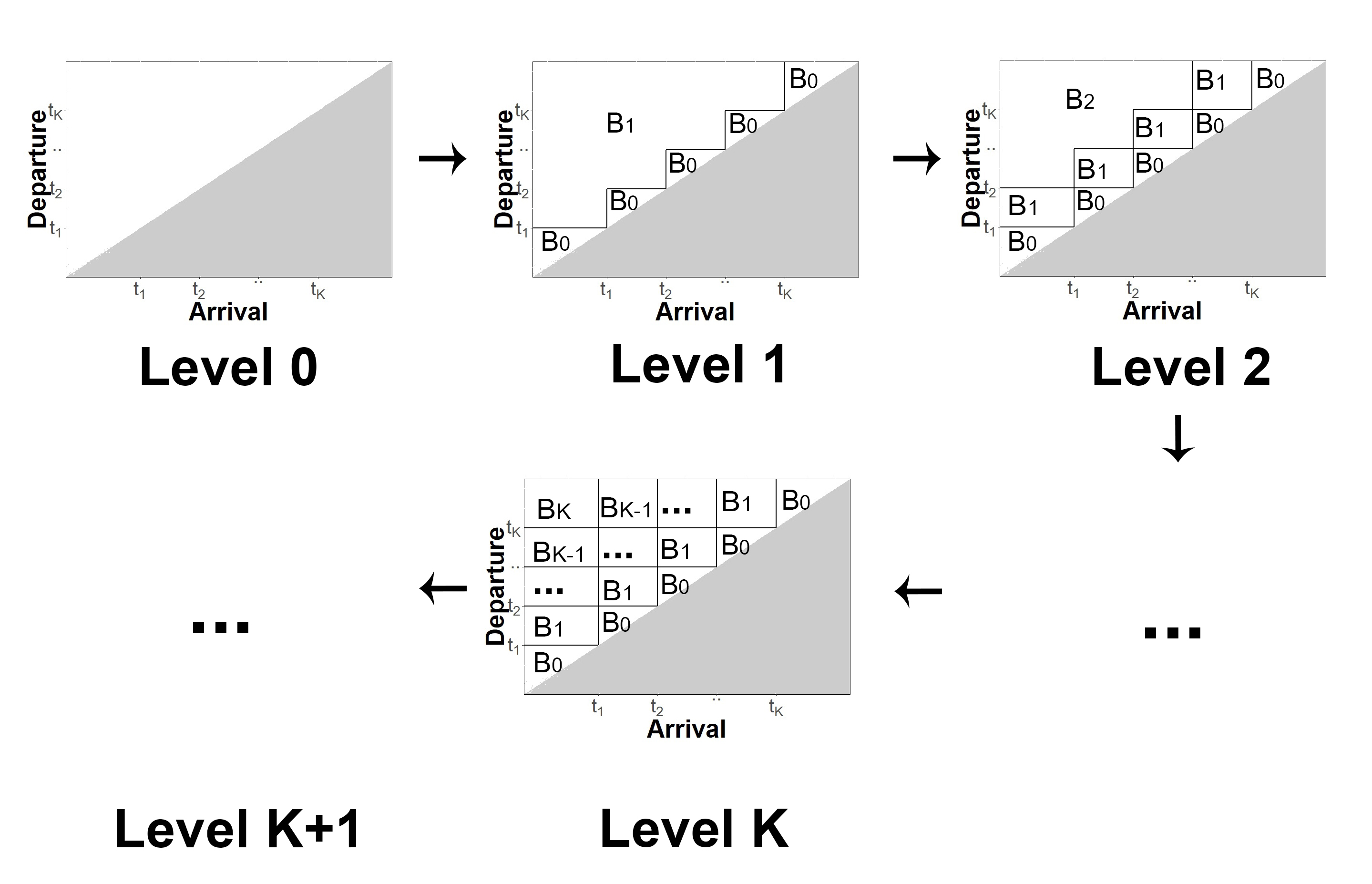}
  \caption{Partition of a PT prior for a distribution defined on $\{ (x,y): (x,y) \in \mathbb{R}^2 , y > x \}$ built using BivLP. The first $K$ levels are built splitting according to the LOS, while the next splits are performed according to the entry dimension.}
\label{fig:3}
\end{figure}

First, we split the sample space in the set $B_0 = \{ (x, y) : t_k < x < y < t_{k+1} \}$, consisting of the individuals departing immediately after arrival and never becoming available for capture, and the set $B_1 = \{ (x, y) : t_k < x < t_{k+1} < y  \}$, consisting of the individuals staying for one or more sampling occasions (level $1$ of Fig. \ref{fig:3}). Next, we split $B_1$ in the set $B_{10} = \{ (x, y) : t_k < x < t_{k+1} < y < t_{k+2} \}$ of individuals staying for only one sampling occasion and the set $B_1 = \{ (x, y) : t_k < x < t_{k+1},  y > t_{k+2}  \}$ of individuals remaining for more than one sampling occasions (level $2$ of Fig. \ref{fig:3}). The process is repeated $K$ times for the remaining sampling occasions. Next, each set is split according to a uniform split using the entry intervals.

\subsection{Figures}

\begin{figure}[h]
\hspace{-0.2in}
  \includegraphics[width=16cm]{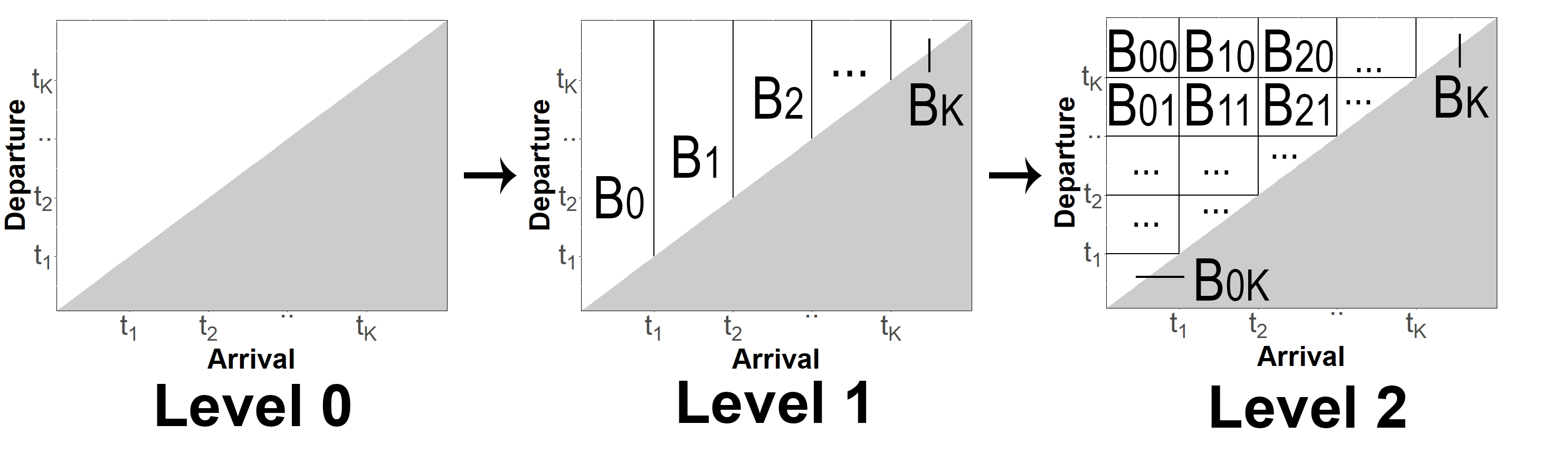}
  \caption{Partition of a PT prior for a distribution defined on $\{ (x,y): (x,y) \in \mathbb{R}^2 , y > x \}$ built using the uniform split twice.}
\label{fig:2}
\end{figure}

\begin{figure}[h]
\hspace{-0.2in}
  \includegraphics[scale=.48]{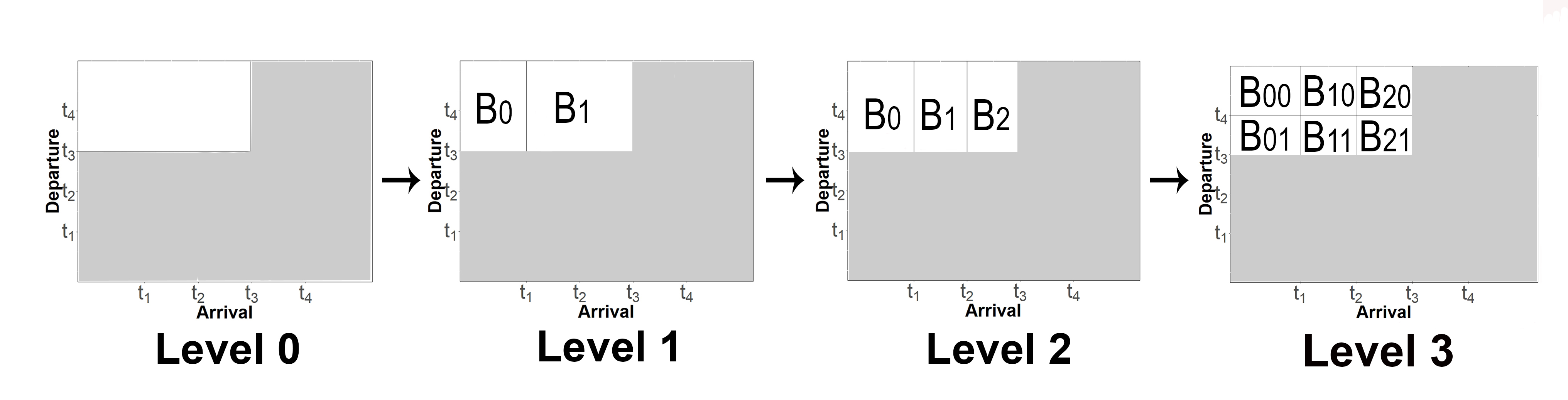}
  \caption{EEBP for the slice $k=3$ of the individuals first captured at time $t_3$.}
\label{fig:truncp}
\end{figure}

\begin{figure}[h]
\hspace{-0.3in}
  \includegraphics[scale = .51]{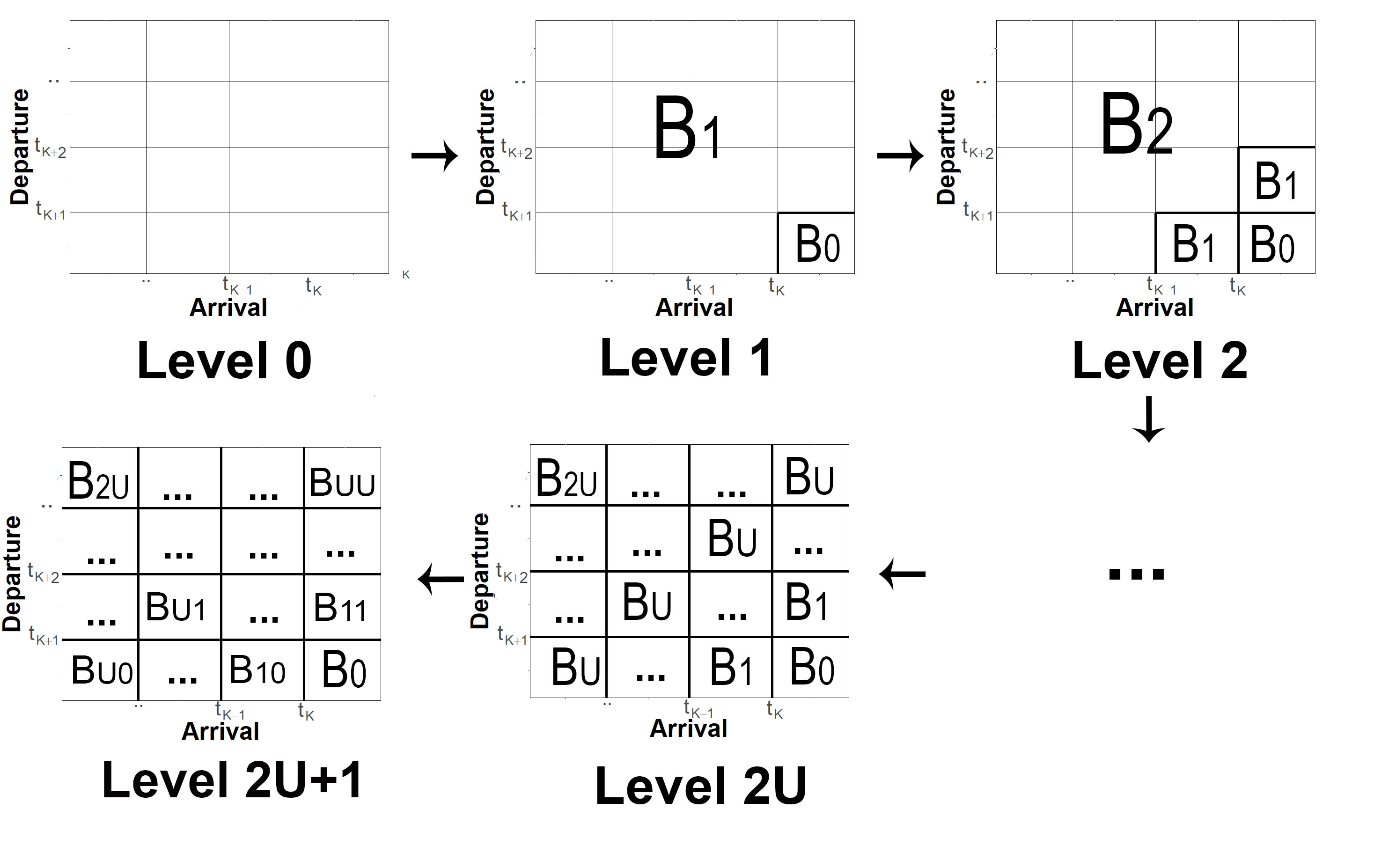}
  \caption{Partition of the PT in the case of RR data for the matrix $\textbf{n}^k$ corresponding to the individuals marked at time $t_k$. The sample space is split in the set of individuals staying for less than one year, $B_0$, and the set of individuals staying for more than one year, $B_1$. After this process is repeated for $2U$ years, each set is split according to the exit intervals.}
\label{fig:rrpt}
\end{figure}

\section{Likelihoods for capture-recapture data}

\subsection{Likelihood of the CJS}
\label{sec:likcjs}

Assuming independence between sampling occasions and between individuals, the probability of observing the CR data, $H$, given the vectors $\textbf{n}^{k}$ can be written explicitly. Although the $\textbf{n}^{k}$ provide the frequencies of all possible presence histories, they do not give us a way to associate individuals and their capture histories to presence histories. For example, an individual with capture history $011$ might have presence history $111$ or $011$. Therefore, we cannot simply express the probability of each observed individual capture history, $H_i$, conditional on the corresponding inferred presence history to construct the likelihood function, but we need to compute the number of possible ways that the capture histories can be assigned to the inferred presence histories. 

Before deriving the likelihood of the general model, we first present a simple case with only $2$ sampling occasions. First, we define as $Z_{ij}$ the number of individual with first capture in sampling occasion $i$ and last capture in sampling occasion $j$. We also recall that $n^k_j$ is the number of individuals with first capture at sampling occasion $j$ and remaining at the site for $j$ sampling occasions. By definition, individuals with capture history $01$ can only be assigned to $n^2_{2}$, from which it follows that the number of possible combinations for assigning the $Z_{22}$ individuals are ${n^{2}_{2}}\choose{Z_{22}}$. The individuals with capture history $11$ can only be assigned to $n^1_{2}$, which gives ${n^1_{2}}\choose{Z_{12}}$ combinations, while the individuals with capture history $10$ can be assigned to either $n^1_{2}$ or $n^1_{1}$, and as $Z_{12}$ are already assigned to $n^1_{2}$, the number of combinations is equal to ${(n^1_{2} - Z_{12}) + n^1_{1}}\choose{Z_{11}}$. If we define as $N_{k}$ and $C_{k}$ the individuals with first capture on sampling occasion $k$ that are available for capture and captured, respectively, the probability of the observed capture histories is


$$
\underbrace{\binom{n_{12}}{Z_{12}} \binom{(n_{12} - Z_{12}) + n_{11}}{Z_{11}} p^{C_1} (1-p)^{N_1 - C_1}}_{\text{individuals captured on the $1^{st}$ sampling occasion}}  \times \underbrace{\binom{n_{22}}{Z_{22}} p^{C_2} (1-p)^{N_2 - C_2}}_{\text{individuals captured on the $2^{nd}$ sampling occasion}}
$$

Using the same rationale for the case with $K$ sampling occasions, the likelihood for the individuals with first capture on sampling occasion $k$, can be expressed as

$$
\hspace{-.5cm}
f_k(H_k | \textbf{n}^k) = \binom{n^k_{K-k}}{Z_{kK}} \binom{(n^k_{K-k} - Z_{k,K}) + n^k_{K-k-1}}{Z_{k,K-1}} \cdots \binom{n^k_{1} + \sum_{l=2}^{K-k} (n^k_{l} - Z_{l+k,K} ) }{Z_{k,k}} p^{C_k} (1 - p)^{N_k - C_k}
$$

where $a$ and $b$ are determined by centering the PT on a given distribution.

\subsection{Likelihood of open-population CR}

We write down the likelihood for the individuals with first capture on sampling occasion $k$, using the same notation of the previous section for the matrix $Z$. 

\begin{itemize}
\item As the individuals in the cell $Z_{kK}$ have to arrive before the $k$-th sampling occasion and depart after the $K$-th, they can only be assigned to the cells $(n^k_{1K},\dots,n^k_{kK})$. Hence, their capture histories can be assigned to the presence histories in ${\sum_{l=1}^k n^k_{lK}}\choose{Z_{kK}}$ possible ways.
\item Similarly, the individuals in cell $Z_{k,K-1}$ have to arrive before the $k$-th sampling occasion and depart after the $K-1$-th and can only be assigned to the cells $(n^k_{1,K-1},\dots,n^k_{k,K-1})$ and to the cells $(n^k_{1K},\dots,n^k_{kK})$ not already assigned to the birds in the cells $Z_{kK}$. Hence, the number of combinations is $\left( \begin{matrix}
\left( \left[ \sum_{l=1}^k n^k_{lK} \right] - Z_{kK} \right) + \sum_{l=1}^k n^k_{l,K-1} \\
Z_{k,K-1}
\end{matrix} \right)$
\end{itemize}

Applying the same rationale for each cell, the likelihood can be written as
\begin{multline}
f_k(H | \textbf{n}^k) = \binom{\sum_{l=1}^k n^k_{lK}}{Z_{kK}} \left( \begin{matrix}
\left( \left[ \sum_{l=1}^k n^k_{lK} \right] - Z_{kK} \right) + \sum_{l=1}^k n^k_{l,K-1} \\
Z_{k,K-1}
\end{matrix} \right) \cdots 
\\
\left( \begin{matrix}
 \sum_{i=k+1}^{K} \left( \left[\sum_{l=1}^k n^k_{l,i} \right] - Z_{ki} \right) + \left(\sum_{l=1}^k n^k_{l,k}\right)  \\
Z_{k,k}
\end{matrix} \right)  p^{C_k} (1 - p)^{N_k - C_k}  
\end{multline}

\section{Long-time series of counts}

\subsection{Partition}

Once the set of splitting rules is introduced, we define the partition of sample space $\Omega_{i_0}$ by first splitting $\Omega_{i_0}$ in $(B^{i_0}_{1},\dots,B^{i_0}_{K_1})$ where $B^{i_0}_{i} = \{ (x,y) \in \Omega_{i_0} , x \in I^1_i \}$, that is, the sample space is split using the dimension of entry according to the $1$-periods. Next, each $B^{i_0}_{i}$ is split into $(B^{i_0}_{i,1},\dots,B^{i_0}_{i,K_1})$, where $B^{i_0}_{i,j} = \{ (x,y) \in B^{i_0}_{i} , y \in I^1_j \}$, that is, each set is now split using the dimension of exit according to again the $1$-periods. 
Analogously, in the next steps, given two sequences of indexes $\epsilon = (\epsilon_1,\dots,\epsilon_n)$ and $\delta = (\delta_1,\dots,\delta_m)$, each set $B^{i_0}_{\epsilon , \delta}$ is split into

$$
\begin{cases}
(B^{i_0}_{\epsilon 1, \delta},\dots,B^{i_0}_{\epsilon K_n , \delta}) \hspace{.1cm} \text{ where }  B^{i_0}_{\epsilon i, \delta} = \{ (x,y) \in  B^{i_0}_{\epsilon, \delta}, x \in I^{n}_{i} \} \hspace{.25cm} \text{ if } n = m \\
(B^{i_0}_{\epsilon, \delta 1},\dots,B^{i_0}_{\epsilon , \delta K_m}) \hspace{.05cm} \text{ where }  B^{i_0}_{\epsilon, \delta i} = \{ (x,y) \in  B^{i_0}_{\epsilon, \delta}, y \in I^{m}_{i} \} \hspace{.2cm} \text{ if } n = m + 1
\end{cases}
$$

\subsection{Coefficients}

We denote by $\beta^{i_0}_{\epsilon, \delta}$ the r.v.s assigning the masses in each level of the partition, that is, $G_{i_0}(\beta^{i_0}_{\epsilon i, \delta}) = \frac{e^{\beta^{i_0}_{\epsilon i, \delta}}}{1 + \sum_{k=0}^{K-2} e^{\beta^{i_0}_{\epsilon k, \delta}}}$ and $G_{i_0}(\beta^{i_0}_{\epsilon, \delta j}) = \frac{e^{\beta^{i_0}_{\epsilon , \delta j}}}{1 + \sum_{k=0}^{K-2} e^{\beta^{i_0}_{\epsilon, \delta k}}}$. We also make the following assumptions on the r.v.s $G_{i_0}$:

\begin{itemize}

\item We assume an HLPT structure across the distributions $G_{i_0}$. This implies that the r.v.s $\beta^{i_0}_{\epsilon, \delta}$ have a common prior $\beta^{i_0}_{\epsilon, \delta} \sim \text{N}(\mu_{\epsilon, \delta}, \sigma^2)$.

\item We assume that $\beta_{\epsilon i, \delta j}^{i_0}$ is the same across the different entry periods, {\it i.e.} $\beta^{i_0}_{\epsilon i, \delta j} = \gamma_{\epsilon, \delta j}^{i_0} \forall i$. 


\item We assume that the masses in periods that are in close proximity are more strongly correlated than the mass of periods farther away from each other, hence we assume that the means $(\mu_{\epsilon 1, \delta},\dots,\mu_{\epsilon K, \delta})$ of the common prior are drawn jointly from a Gaussian process $\text{GP}((\mu^0_{\epsilon 1, \delta},\dots,\mu^0_{\epsilon K, \delta}), \sigma^2_0 k_l)$, where $k_l$ is the Gaussian covariance function $k(x_i,x_j) = \exp(-\frac{(x_i - x_j)^2}{l^2}) $ with parameter $l$.

\end{itemize}

As a result, $\sigma^2$ tunes the variability across $0$-periods, $\sigma^2_0$ tunes the variability across $i$-periods and $l$ is the correlation across $i$-periods.

\section{Case studies}

\subsection{Spoonbills}

\subsubsection{Model specification}

As opposed to the model described in Section $5.2$ of the paper, we have two different populations of individuals, the ones already marked and the unmarked, and hence we have to modify the model that has been introduced in the previous section. Similarly to the case with one population, we model the marked individuals resighted using the variables $(t^{i}_1,t^{i}_2)$, while the marked individuals never resighted and the unmarked individuals are summarised in two matrices $\{ n^{M}_{fl} \}$ and $\{ n^{U}_{fl} \}$, respectively, where cell $(f,l)$ corresponds to the number of individuals first available for capture at time $t_f$ and last available at time $t_l$.

The number of marked birds, $N_M$, and the number of unmarked birds, $N_U$, are assigned two Poisson prior distributions, $N_M \sim \text{Poisson}(\omega_M)$ and $N_U \sim \text{Poisson}(\omega_U)$, respectively. We assume that each marked individual can be resighted by visiting its nest with probability $p_C$ and through camera traps with probability $p_R$. As the unmarked birds are also detected through camera traps, we assume they can be detected with the same probability $p_R$ of resighting an already marked bird. We note that this is a realistic, but also necessary assumption in this case, as otherwise the CD do not include enough information to separately estimate the population size $N_U$ of unmarked birds and the probability of resighting an unmarked bird, $p_R$, if that is different to marked birds.  The MCMC scheme are discussed in Section $6.2$ of the supplementary material.


\subsubsection{Prior specification}

We choose a uniform prior distribution for the resighting probabilities and a Gamma prior distribution with mean $50$ and variance $4000$ for the intensities $\omega_M$ and $\omega_U$ of the two population sizes. 

Inference is performed using an MCMC algorithm by sampling from the posterior distribution of the individual entry and exit intervals $(t^{i}_1,t^{i}_2)$, the two matrices $\{ n^{M}_{fl} \}$ and $\{ n^{U}_{fl} \}$, as well as the probabilities $\omega_{ij}$ of the PT and the resighting probabilities. The MCMC is described in Section $6.2$.

\subsubsection{Figures}

\begin{figure}[H]
\begin{subfigure}{.5\textwidth}
  \centering
   \includegraphics[scale = .5]{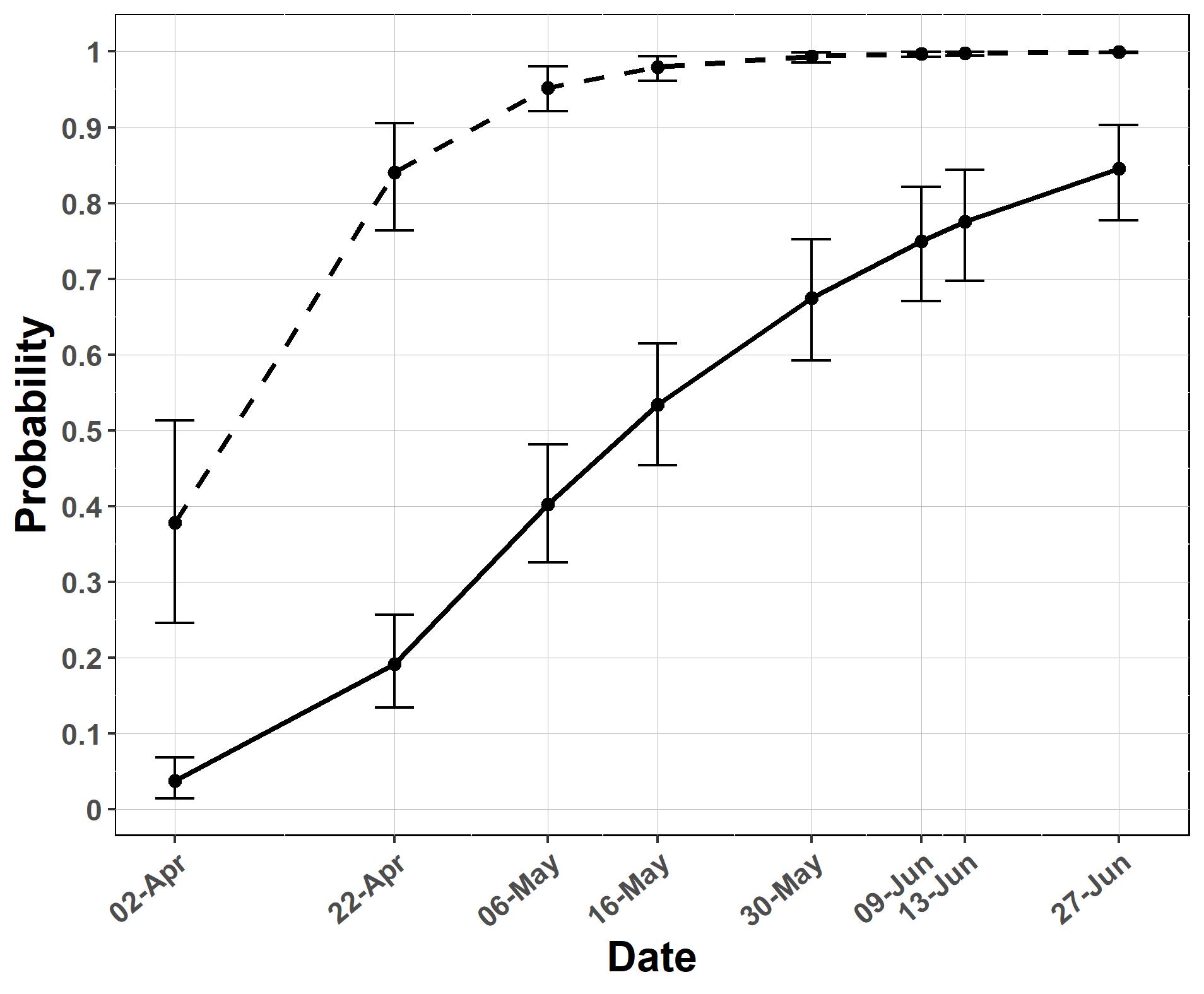}
  \caption{}
\end{subfigure}%
\begin{subfigure}{.5\textwidth}
  \centering
  \begin{tabular}{|l|l|l|l|}
\hline
      & 2.5\% & Mean & 97.5\%  \\ \hline
$N_M$ & $39$   & $48$   & $59$ \\ \hline
$N_U$ & $45$   & $65$   & $91$  \\ \hline
$p_R$ & $.30$  & $.38$  & $.48$  \\ \hline
$p_C$ & $.21$  & $.29$  & $.38$  \\ \hline
\end{tabular}
  \caption{} 
\end{subfigure}
\caption{(a) Posterior mean of the CDF of entry intervals (dashed line) and exit intervals (solid line), with confidence bars showing the 95\% PCIs. The ticks on the $x$-axis represent the sampling occasions. (b) Posterior summaries of population sizes and detection probabilities.}
  \label{fig:density_spoonbills}
\end{figure}

\subsection{Ring-Recovery}

\subsubsection{Model specification}

We summarise the data in two $K \times K$ matrices, $R^A$ and $R^J$, for adults and juveniles, respectively. The total number of juveniles and adults marked in year $k$, are denoted by $m^J_k$ and $m^A_k$, respectively. The first row of the matrix $\textbf{n}^k$, $\{ n^k_{0 u_l} \}_{ u_l=0,\dots,U}$, consists of the juveniles, and its sum is thus equal to $m^J_k$, while the rows from $2$ to $U + 1$ consist of the individuals marked as adults, and their sum is equal to $m^A_k$. Assuming a Beta prior for the recovery probability, $\lambda$, the model can be summarised as
\begin{equation}
\label{eq:rr2}
\begin{cases}
R^{y}_{k,k+l} \sim \text{Binomial}(n^k_{0 u_l}, \lambda)  \hspace{3cm} 
R^{a}_{k,k+l} \sim \text{Binomial}(\sum_{u_f=1}^U n^k_{u_f u_l}, \lambda)\\
n^k_{0 u_l} \sim \text{Multinomial}(m_k^J, \omega_{0 u_l}) 
\hspace{2.4cm} 
\{ n^k_{u_f u_l} \}_{f > 0} \sim \text{Multinomial}(m_k^A, \omega_{u_f u_l}) \\
\{ \omega_{u_f u_l} \} \sim \text{PT}( \Pi, \{ \alpha \})  
\hspace{3.7cm}
\lambda \sim \text{Beta}(a_0, b_0) 
\end{cases}
\end{equation}
The first row of matrix $n^k_{u_f u_l}$ contributes to the recoveries $R^y$ for the juveniles and the rest of the matrix contributes to the recoveries $R^a$ of the adults.

We perform inference by updating the random matrices $n^k_{u_f u_l}$, the probabilities $\omega_{u_f u_l}$ and the recovery probability $\lambda$. Details of the MCMC sampler are presented in Section $6.3$.

\subsubsection{Prior specification}

Instead of assuming a centering distribution on the tree, we assume uniform prior distributions directly on the parameters of the Beta distribution assigning the masses in each split of the PT. Moreover, we assume a Beta$(1,1)$ prior distribution for the recovery probability $\lambda$ and we set the upper bound $U$ on the age of the individuals to $18$. 

\subsubsection{Figures}

\begin{center}
\begin{figure}[H]
\hspace{1cm}
\includegraphics[scale=.5]{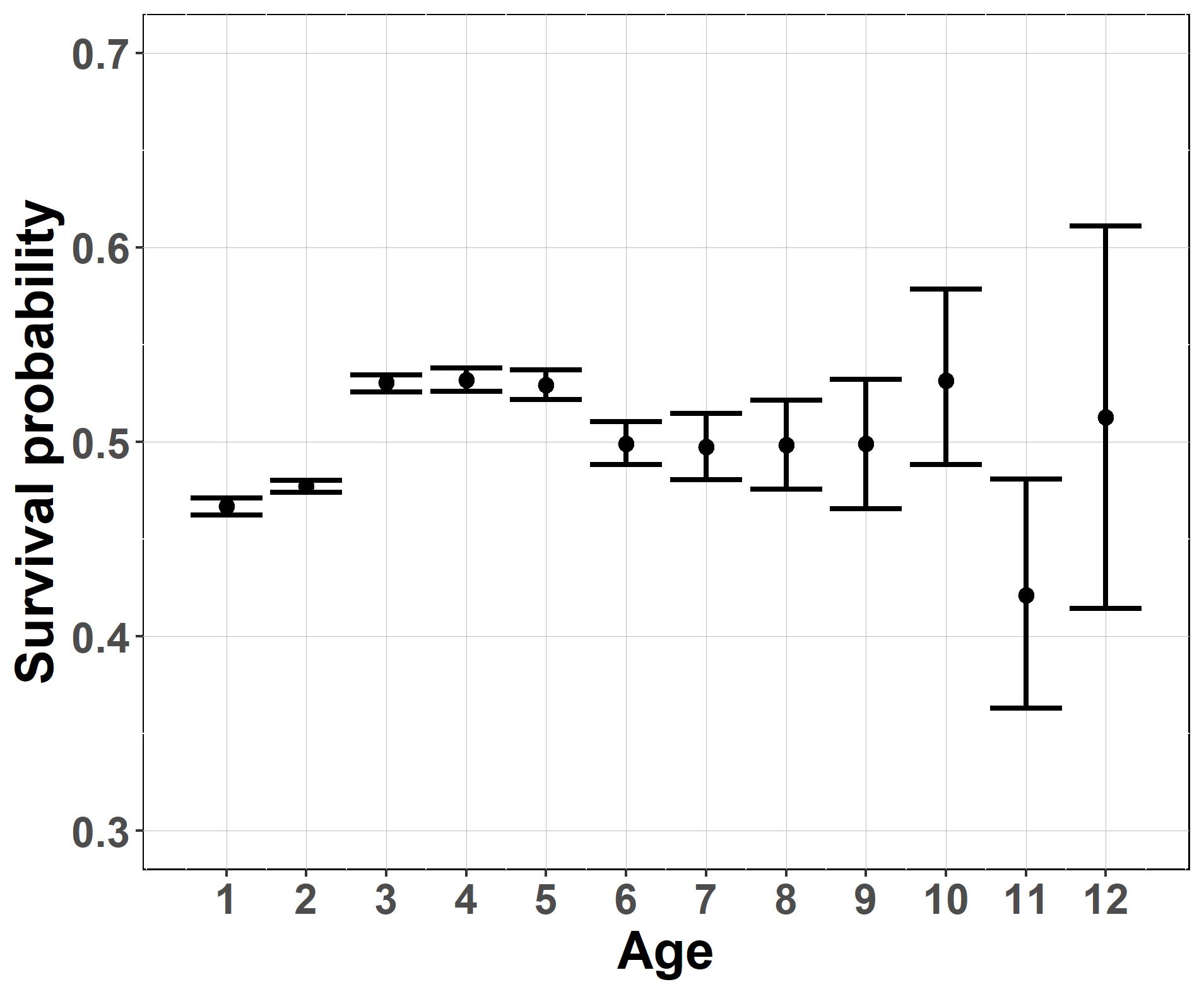}
\caption{Case study on mallards. Posterior means of the probabilities $\phi_j$ of surviving to age $j+1$ conditional on being alive at age $j$.}
\label{fig:survival}
\end{figure}
\end{center}

\subsection{Newts}

\subsubsection{Prior specification}

We center the HLPT on a bivariate distribution with double exponential marginal distributions, with parameters $\mu = (\mu_1, \mu_2)$ and $\lambda = (\lambda_1, \lambda_2)$. Prior information on the entry and exit distribution is elicited through the prior on $\eta = (\mu, \lambda)$. Prior knowledge suggests that a considerable number of individuals tend to enter the site between the beginning of March and the end of April and exit between the end of May and the end of July, and hence we choose hyperpriors for $\eta = (\mu, \lambda)$ such that $95 \%$ of the prior mass of the entry and exit density is in the aforementioned ranges. 

\subsubsection{Figures}

\begin{figure}[H]
\begin{tabular}{lll}
\includegraphics[scale=.5]{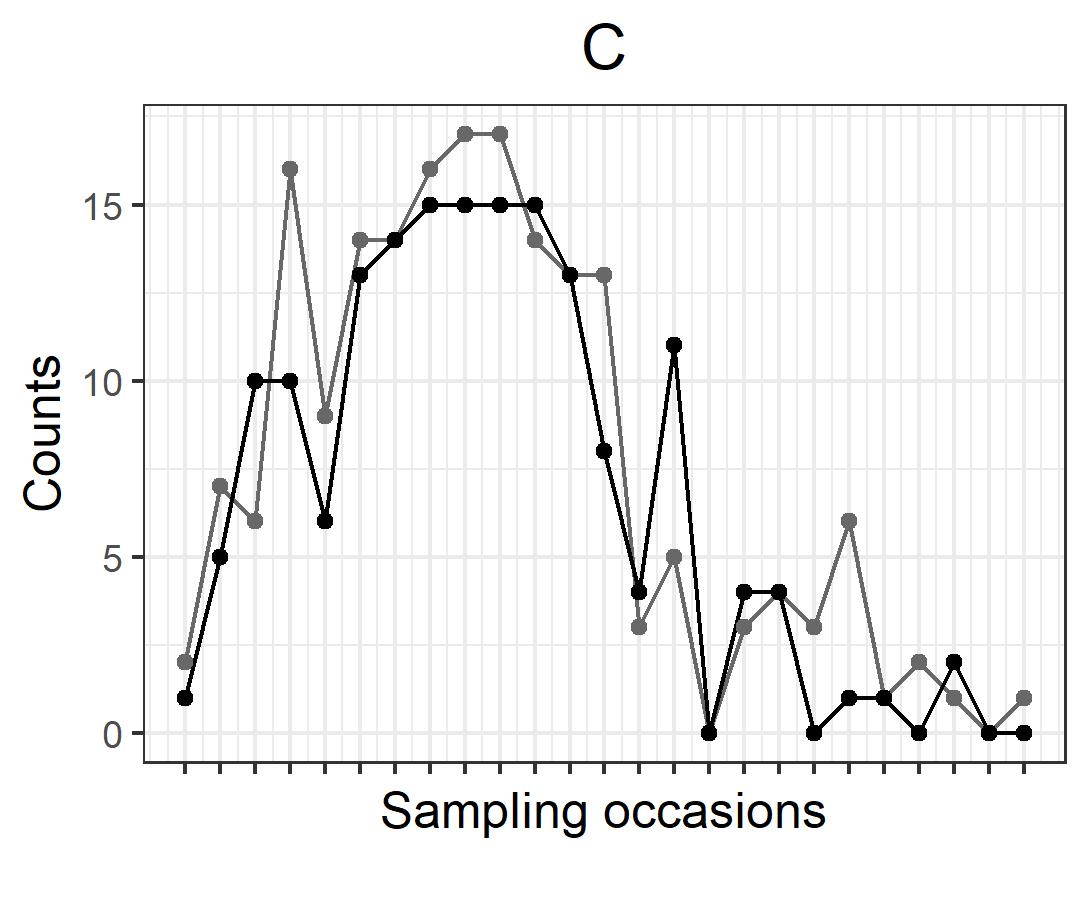}
 & 
 \includegraphics[scale=.5]{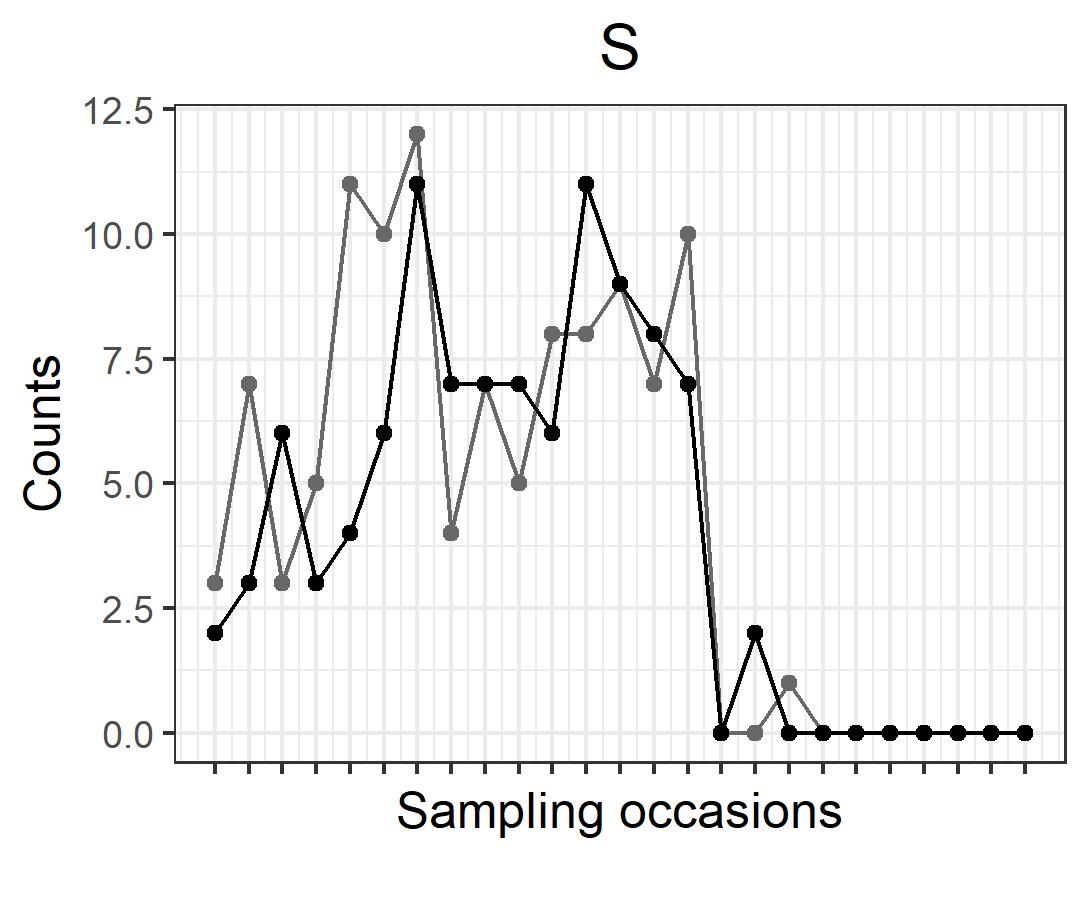}
  & 
 \includegraphics[scale=.5]{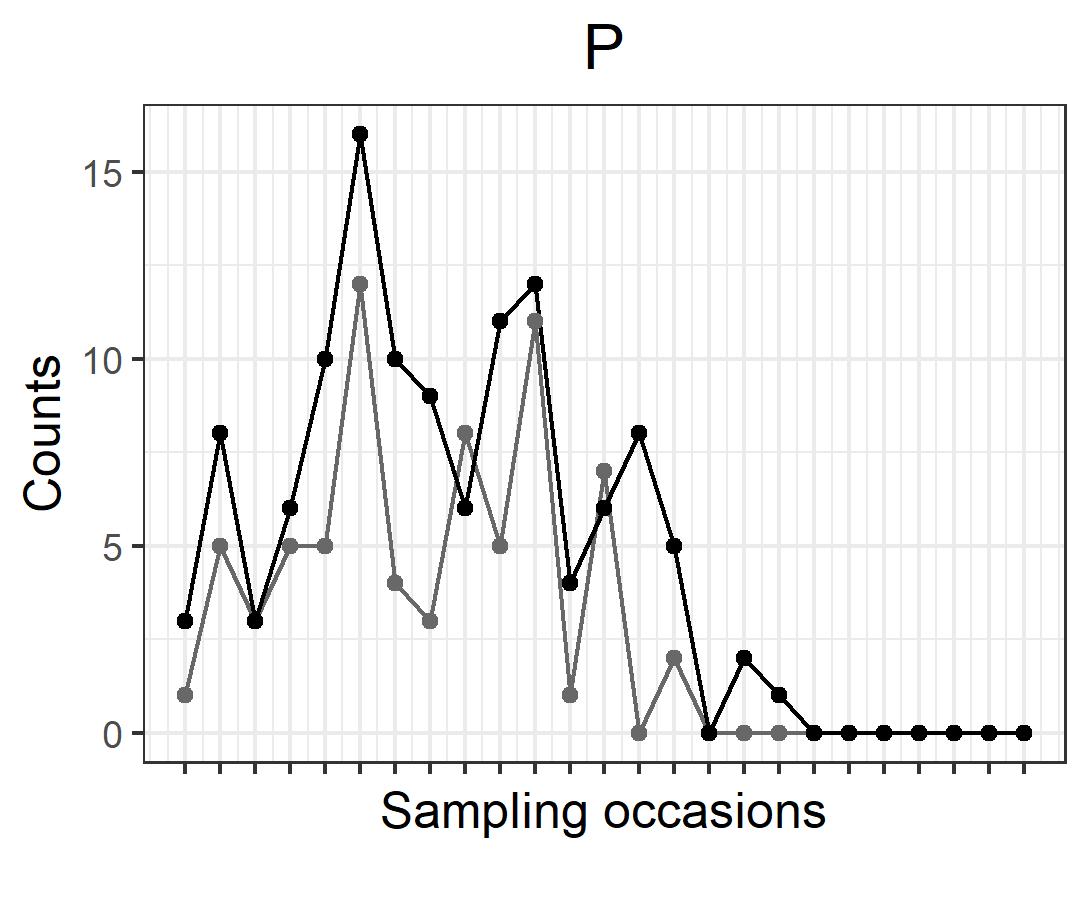}
\end{tabular}
\caption{Number of newts detected on each sampling occasion shown for great crested newts, smooth and palmate newts, respectively. In each plot, the male are represented in grey and the female in black.}
\label{fig:newts}
\end{figure}

\begin{figure}[H]
  \centering
  \begin{minipage}[b]{0.32\textwidth}
    \includegraphics[scale=.4]{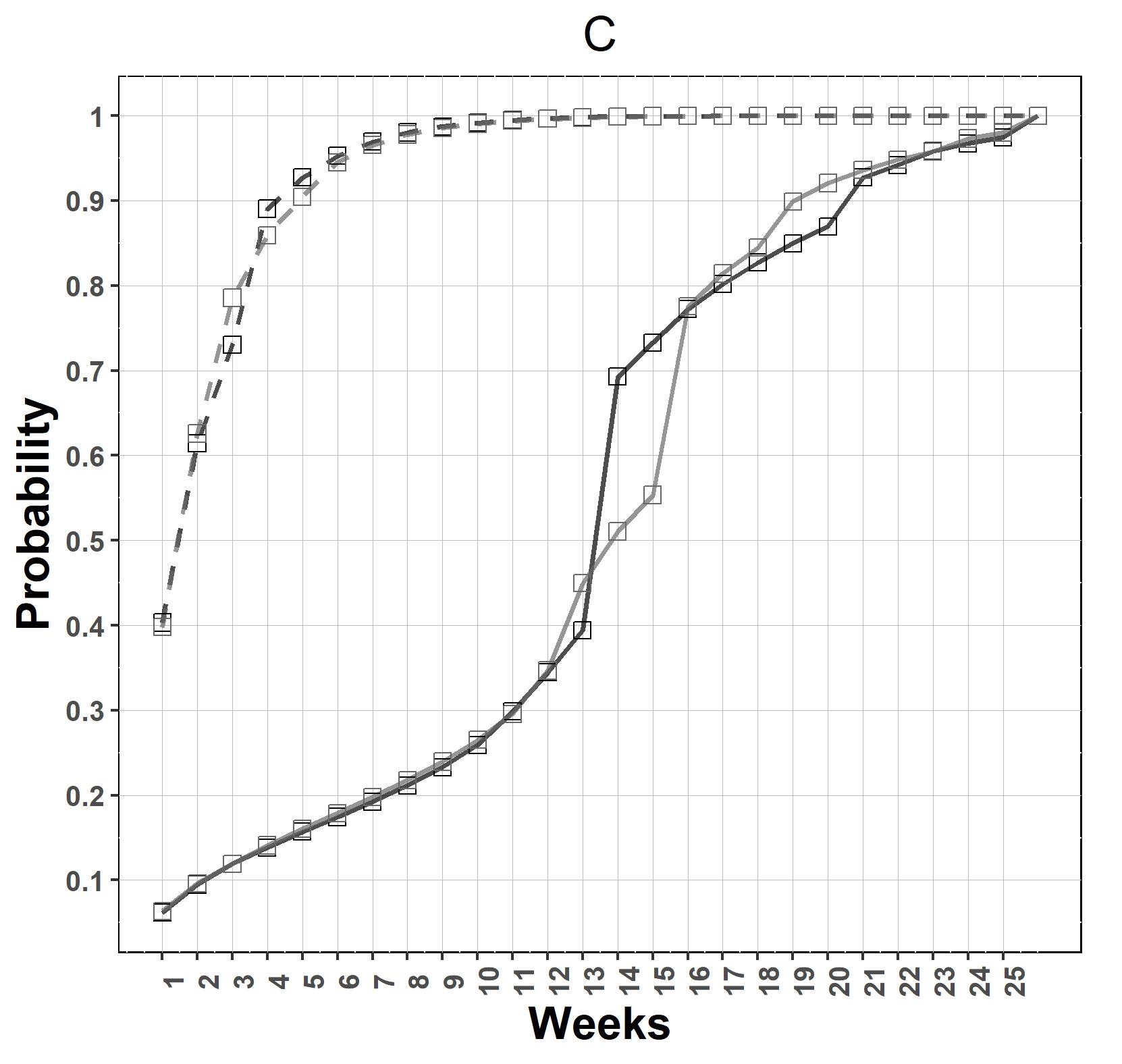}
  \end{minipage}
  \hfill
    \begin{minipage}[b]{0.32\textwidth}
    \includegraphics[scale=.4]{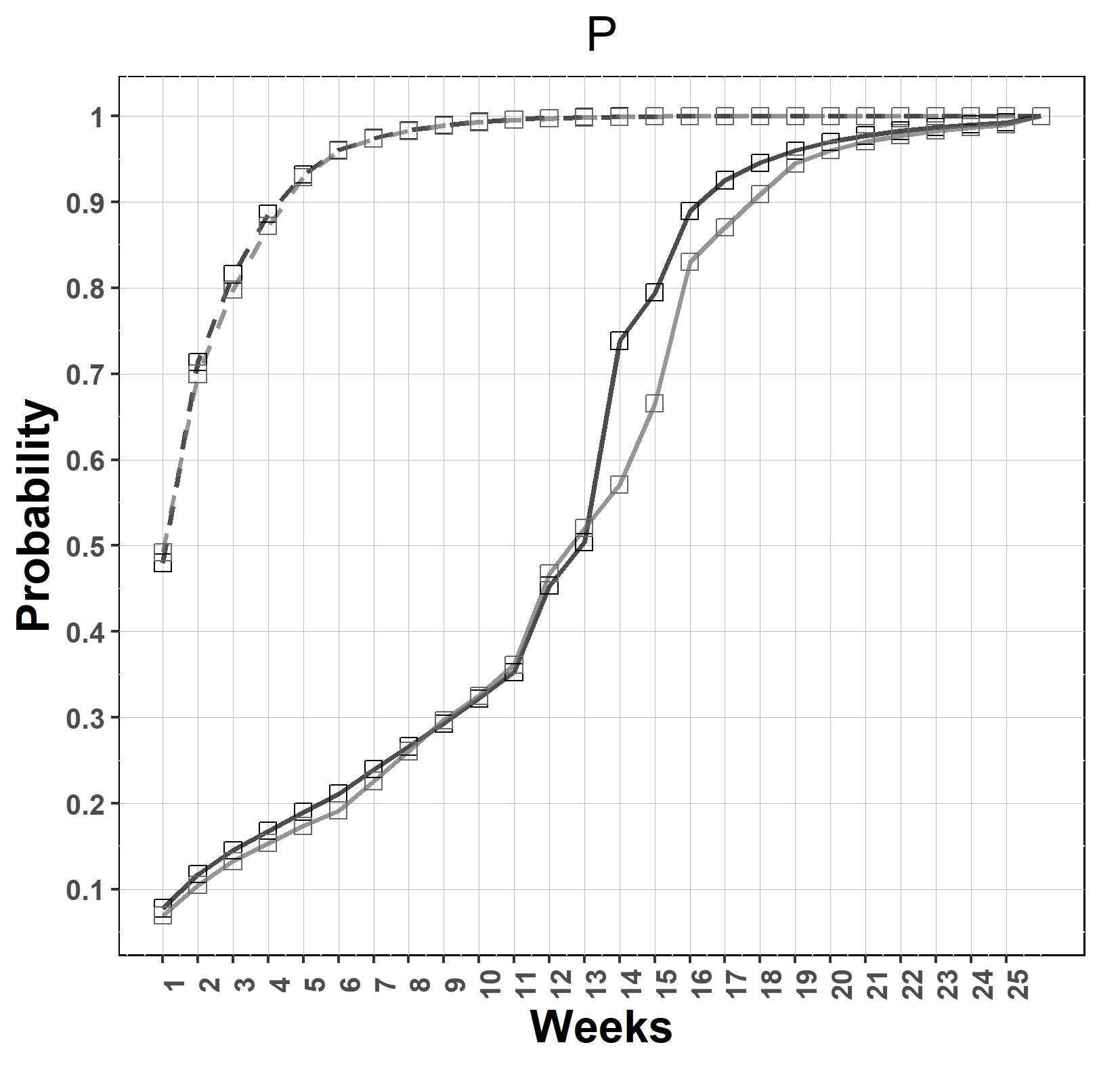}
  \end{minipage}
\hfill
  \begin{minipage}[b]{0.32\textwidth}
    \includegraphics[scale = .4]{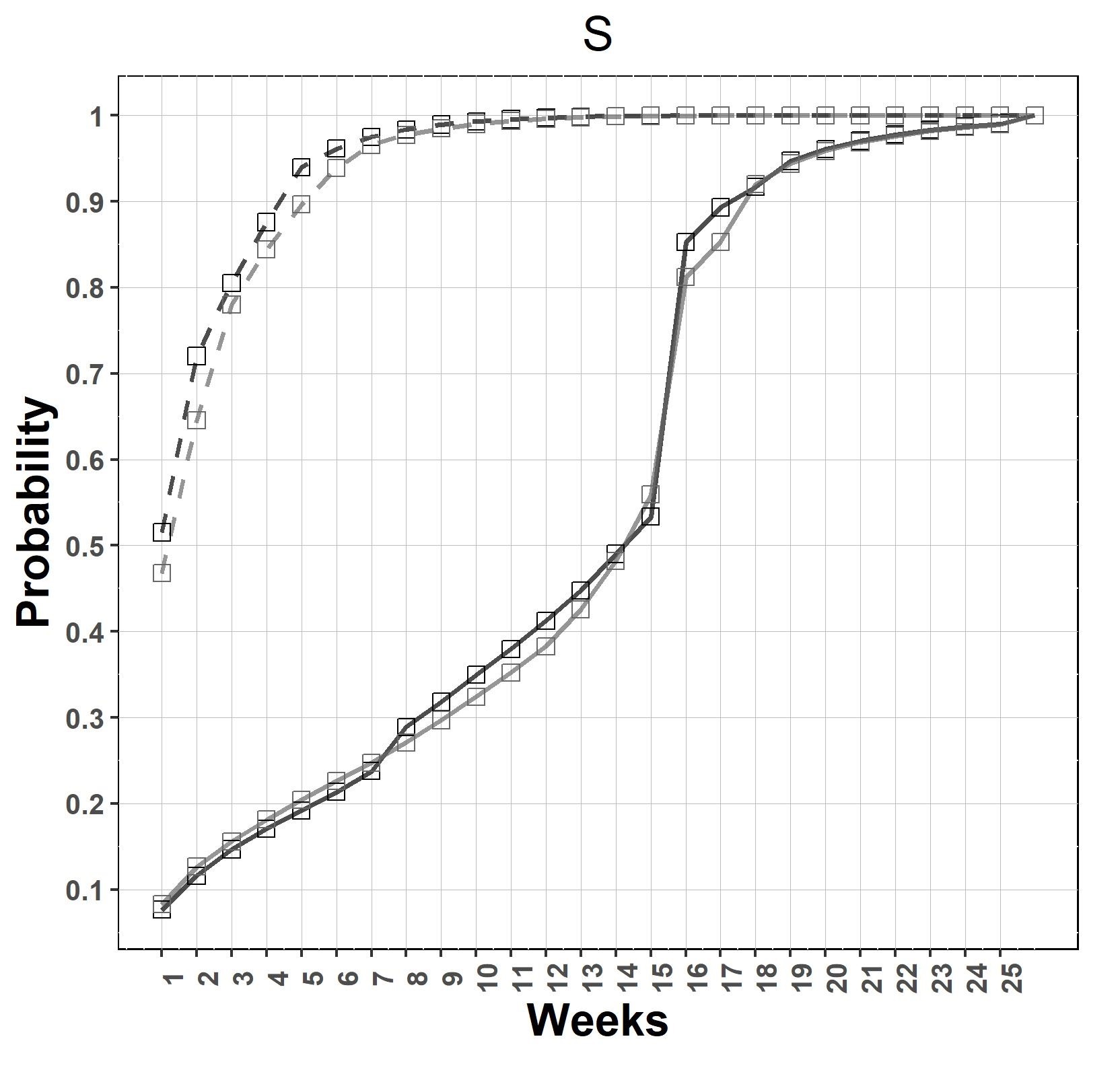}
  \end{minipage}
      \caption{Posterior CDFs of the entry (dotted line) and exit (solid line) distribution for male (black) and female (grey) great crested (C), palmate (P) and smooth (S) newts. }
     \label{fig:densities_newts}
\end{figure}

\begin{figure}[H]
\hspace{-0.2in}
  \includegraphics[width=17.5cm]{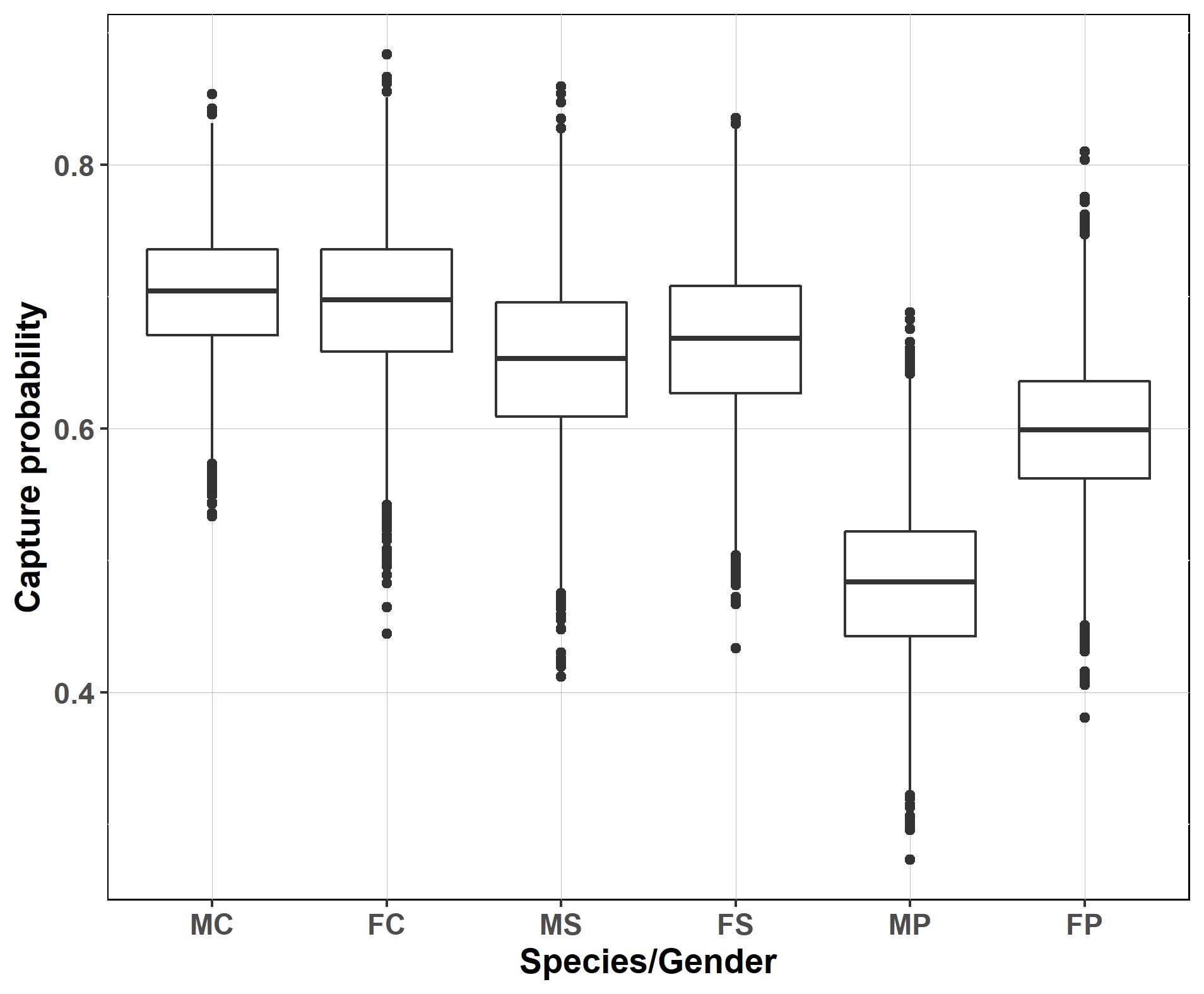}
  \caption{Boxplots of the capture probabilities for the different species and genders.}
\label{fig:prob_newts}
\end{figure} 

\subsection{Moths}

\subsubsection{Prior specifications}

We center the PT for the entry and exit distribution on two normal distributions with parameter $\mu = 2$ and $11$, respectively, and variance $\lambda = 4$, where the unit is taken to be a month. We take a uniform prior distribution for the detection probability $p$ and for the mean $\omega$ of the population size we take a Gamma$(40, .2)$, where the parameters have been chosen so that the mean of the distribution is $200$ and the variance is $1000$. Moreover, we take $\sigma_0 = \sigma = l = 1$ and the prior $\rho$ on the probability of stopping equal to $0.1$. 

\subsubsection{Figures}

\begin{figure}[H]
\hspace{0.0in}
  \includegraphics[scale = .45]{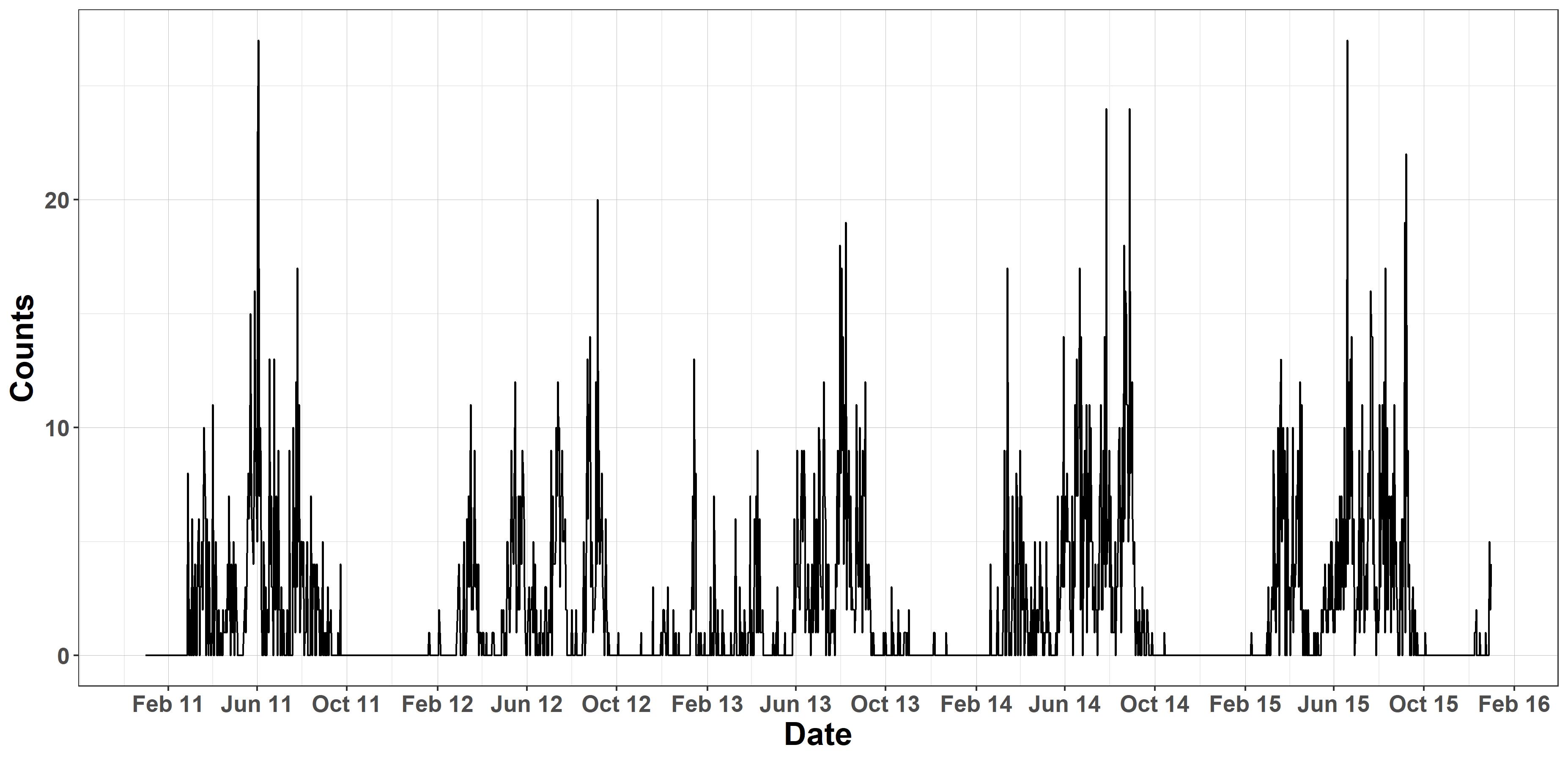}
  \caption{Daily counts of moths detected at the site.}
\label{fig:moths}
\end{figure} 

\begin{table}[H]
\hspace{3cm}
\begin{tabular}{|l|l|l|l|l|l|}
\hline
Year & $2011$ & $2012$ & $2013$ & $2014$ & $2015$ \\ \hline
Temperature & $27.14$  & $35.28$   & $42.98$   & $27.42$    & $30.21$  \\ \hline
\end{tabular}
\caption{Average daily minimum temperature from the $1$st to the $14$th of January for each year, measured in Farenheit.}
\label{table:moths_temp}
\end{table}

\begin{figure}[H]
\hspace{.5in}
  \includegraphics[scale=.55]{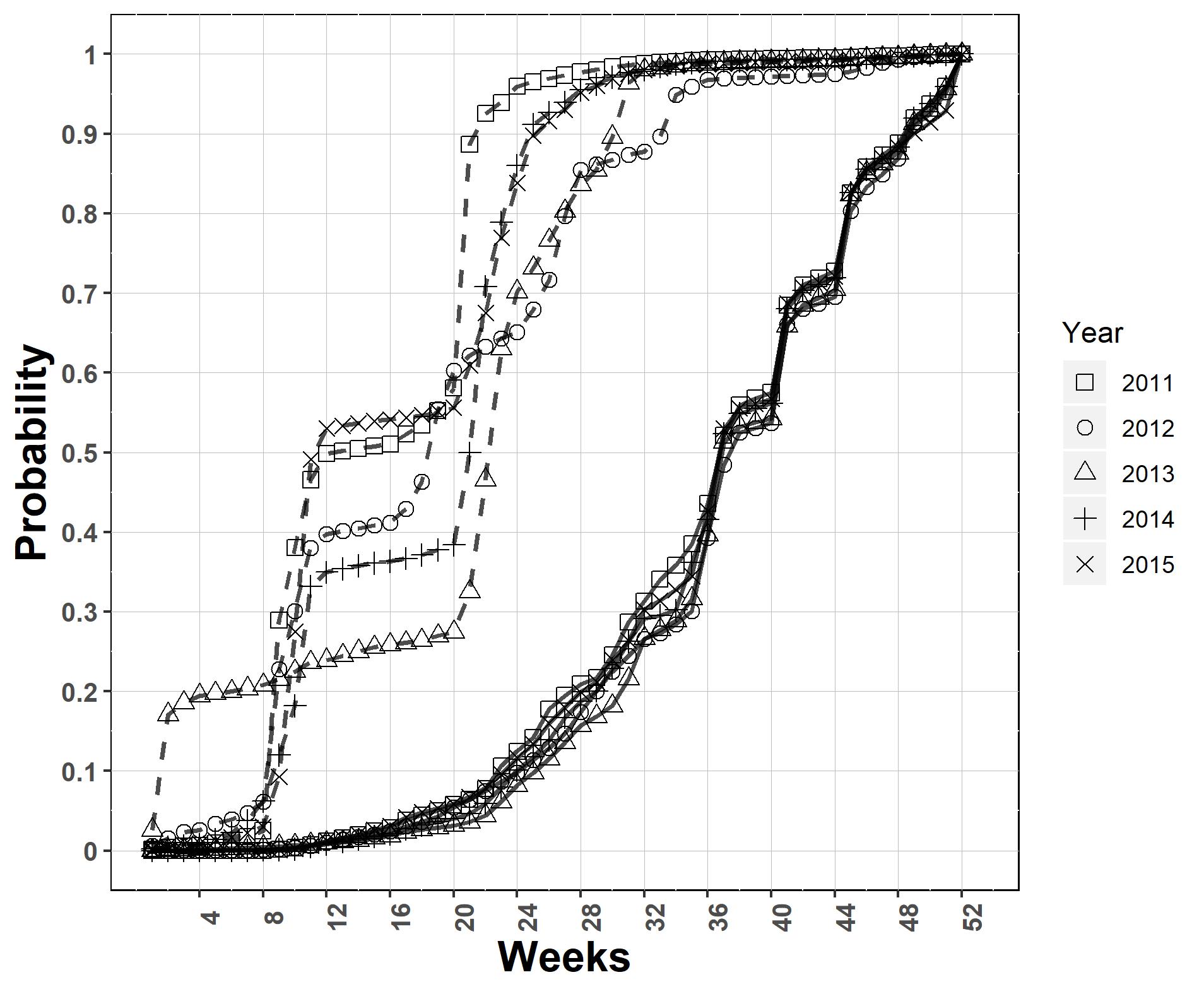}
  \caption{Case study on moths.  Posterior mean of the entry (dashed line) and exit (solid line) CDFs. Each year is represented by a different symbol.}
\label{fig:density_moths}
\end{figure}

\section{Posterior inference}

\subsection{Cormack-Jolly-Seber model}
\label{mcmc0}

The parameters to be updated are the latent number of individuals $\{ n^k_{j} \}$, the random variables $V^k_j$ and the capture probability $p$. 

To ease the computation, we introduce the probabilities $\omega^k_j$ of an individual with first capture at time $t_k$ to have the time of possible last capture at time $t_j$. These probabilities can be computed straightforwardly from the $V^k_j$.

We recall that the capture histories of each individual are summarised in a matrix $H$, where $H_{ik}$ is $1$ if the $i$-th individual was captured on sampling occasion $k$ and $0$ otherwise.

\begin{itemize}

\item Update $\{ n^k_{j} \}$

We update each element of the vector with a Metropolis-Hastings step. Recalling that the number of individual with first capture at time $t_k$, $F_k$, is fixed, the posterior distribution given the data $H$ can be written as

$$
p( n^k_{j} | H, \omega^k_{j}, p) \propto p(n^k_{j} | F_k, \omega^k_{j}) p( H | n^k_{j}, p) \propto
$$
$$
\left( \prod_{j} \text{Multinomial}(n^k_{j} | F_k, \omega^k_j))  \right)
f_k(H_k | \{ n^k_{j} \})
$$

where $f_k(H_k | \{ n^k_{j} \})$ is shown in Section \ref{sec:likcjs}.

\item Update $V^k_j$

These are the r.v.s assigning the masses in a PT, and therefore they can be sampled from the posterior distribution of a PT.

For example, in the unconstrained case, the full conditional is

$$
V^k_j \sim \text{Beta}\left( a + n^k_{j} , b + \sum_{l=j+1}^{K-k+1} n^k_l \right)
$$

\item Update  $p$ 

This parameter is sampled straightforwardly from its full conditional:
$$
p \sim \text{Beta}\left(1 + \sum_{i=1}^D \sum_{j=1}^K H^{ik} - \tilde{N} , 1 + \sum_{j=1}^K N_j - \left( \sum_{i=1}^D \sum_{j=1}^K H^{ik} - \tilde{N} \right)  \right)
$$
where $N_j$ is the number of individual available for capture (after their first) at sampling occasion $j$ and $N$ is the total number of individuals available for capture at least once  after their first capture.

\end{itemize}

\subsection{Joint model for capture-recapture and count data}
\label{mcmc1}

The captures are arranged in the matrices $H^1$ and $H^2$, respectively. The parameters to be updated are the entry and exit intervals of the already marked and captured individuals $(t_1^i,t_2^i)$, the latent number of not resighted already marked individuals $\{n_{ij}^M\}$, the latent number of unmarked individuals $\{n_{ij}^U\}$, the detection and resightings probabilities $p_{R}$ and $p_C$ and the probabilities $\omega_{ij}$ sampled from the PT. 

\begin{itemize}

\item Update $(t_1^i,t_2^i)$

We update these variable with a random walk MH algorithm, where we for each individual we propose a new elemement $( (t_1^i)^{\star} , (t_2^i))^{\star} )$ from $\{ t_1^i - 1, t_1^i, t_1^i + 1 \} \times \{ t_2^i - 1, t_2^i, t_2^i + 1 \}$ . 

The posterior distribution can be written as

$$
p(t_1^i,t_2^i | H^1_i , H^2_i,  \{ D_k \}, \{ \omega_{ij}), p^R, p^C \}) \propto 
$$
$$
p(H^1_i | t_1^i,t_2^i | p_R) \ p(H^2_i | t_1^i,t_2^i | p_C)\ \left( \prod_{k=1}^K p(C_k | N_k, p_R) \right) \ p( t_1^i,t_2^i | \{ \omega_{ij}) \} ) 
$$
where $p(H^1_i | t_1^i,t_2^i | p) = p^{\sum_{k=1}^K H^1_{ik}} (1 - p)^{t_2^i - t_1^i + 1 - \sum_{k=1}^K H^1_{ik}}$.


\item Update $\{ n^M_{ij} \}$ and $\{ n^U_{ij} \}$ 

These variables are updated with a MH algorithm, where for each cell the number of elements we propose to add/remove is sampled from a geometric distribution.

The posterior distribution can be written similarly to the $(t_1^i,t_2^i)$.

\item Update $\omega_{ij}$.

If we summarise all the individuals in a matrix $n_{ij}$, where cell $(ij)$ has the number of individuals with first possible capture at time $t_i$ and last possible capture at time $t_j$, the probabilities $\omega_{ij}$ can be sampled from the posterior of a PT given the variables $n_{ij}$.

\item Update $p_R$ and $p_C$

These probabilities can be sampled directly from their full conditionals: 

$$
p^R \sim \text{Beta}\left( 1 +  \sum_{j=1}^K \sum_{i=1}^D H^1_{ij} + \sum_{j=1}^K D_j, 1 + \sum_{j=1}^K N_j^M - \sum_{j=1}^K \sum_{i=1}^D H^1_{ij} + \sum_{j=1}^K N_j^U - \sum_{j=1}^K D_j \right)
$$
$$
p^C \sim \text{Beta}\left( 1 +  \sum_{j=1}^K \sum_{i=1}^D H^2_{ij}, 1 + \sum_{j=1}^K N_j^M - \sum_{j=1}^K \sum_{i=1}^D H^2_{ij} \right)
$$
where $N_j^M$ and $N_j^U$ are the number of marked and unmarked individuals, respectively, available for capture in sampling occasion $k$.

%
%
%

\end{itemize}

\subsection{Ring Recovery model}
\label{mcmc4}

The parameters to be updated are the elements $\{ n^k_{ij} \}$ in the entry and exit matrix, the probabilities $\{ \omega_{ij} \}$ and the recovery probabilities $p_R$.

\begin{itemize}

\item Update $\{ n^k_{ij} \}$

These variables can be updated with a MH algorithm. The posterior can be written as

$$
p(n^k_{ij} | H^A, H^J, \omega_{ij}, m^A_k, m^J_k, p_R) \propto p( n^k_{ij} | m_k, \omega_{ij}) \ p(H^J_{kj} | n^k_{ij}, \lambda)
$$

where the first term is a multinomial distribution and the second term is a product of binomial distributions.

\item Update $\{ \omega_{ij} \}$

These parameter are sampled using the standard update of the PT. As we use a RPT across the different matrices $\textbf{n}^k$, in order to use the standard update of the PT we summarise all the individuals in a single matrix $n_{ij}$ by collapsing the set of matrices $n^k_{ij}$ across the dimension $k$.

\item Update $\mu$ and $\sigma$

It is useful, in order to write down the posterior distribution of these parameters, to introduce the variables $n_{\epsilon}$, representing the number of elements belonging to the set $B_{\epsilon}$ of the partition. The marginal likelihood of $n_{\epsilon}$ can be expressed as

$$
p(\{ n_{\epsilon} \} | \{ \alpha_{\epsilon} \}) = \prod_{\epsilon} \int \textnormal{Bin}(n_{\epsilon0} | n_{\epsilon}, q_{\epsilon0}) \textnormal{Beta}(q_{\epsilon0} | \alpha_{\epsilon0}, {\alpha}_{\epsilon1} )  dq_{\epsilon0}.
$$
Integrating out the probabilities $q_{\epsilon0}$ gives as a result a beta-binomial distribution, from which it follows that the marginal likelihood of the latent variable $n_{\epsilon}$ given the hyperparameter $\eta$ is

$$
p(\left \{n_{\epsilon} \right \} | \eta) \propto \prod_{\epsilon} \frac{B(\alpha_{\epsilon 0} + n_{\epsilon 0}, \alpha_{\epsilon 1} + n_{\epsilon 1})}{ B(\alpha_{\epsilon 0}, \alpha_{\epsilon 1})}.
$$

Given the marginal likelihood, the parameters can be updated using a Metropolis-Hastings random walk update step.

\item Update $\lambda$

The parameter can be sampled easily from the full conditional as the prior is a Beta distribution and the likelihood is a product of Binomial likelihoods.

\end{itemize}

\subsection{Hierarchical model}
\label{mcmc2}

The parameters to be updated are the latent number of individuals $n^s_{ij}$, the parameters $\beta^s_{\epsilon}$ assigning the probabilities of entering and exiting for each site, the hyperparameters $\eta$ and the capture probabilities $p_s$.

To write the posterior of the $\beta$'s and the hyperparameter $\eta$ (or equivalently of the parameters $\mu$ as there is a deterministic relationship between the parameters $\mu$ and $\eta$) given the latent variable $n^s_{ij}$, we work with the variables $\tilde{n}^s_{\epsilon_1,\dots,\epsilon_m}$, representing the number of elements in the $s$-th dataset belonging to the set $B_{\epsilon_1 \dots \epsilon_m}$ of the partition. The posterior can be written as:

$$
\hspace{-1cm}
p(\beta, \eta | \{ \tilde{n} \}) \propto p(\eta) \prod_{s = 1}^S  \text{Pois}(n^s_{\Omega}, \omega_s) \left( \prod_{\epsilon}  \frac{(e^{\beta^s_{\epsilon}})^{n^s_{\epsilon 0}}}{(1 + e^{\beta^s_{\epsilon}})^{n^s_{\epsilon}}}  p(\beta^s_{\epsilon} | \eta)  \right)
$$

where $n^s_{\Omega}$ is the number of elements in the $s$-th dataset. Now as in \citet{polson2013bayesian} we add a variable $\omega_{\epsilon} \sim \text{PG}(\tilde{n}_{\epsilon},0)$ and write the joint posterior distribution of all the variables as:

$$
p(\beta, \eta | \{ \tilde{n} \}, \omega_s) \propto p(\eta) \left( \prod_{s = 1}^S \text{Pois}(n^s_{\Omega}, \omega_s) \prod_{\epsilon}  e^{k^s_{\epsilon}\beta^s_{\epsilon
}} e^{- \frac{\omega^s_{\epsilon}(\beta^s_{\epsilon})^2}{2}} p(\omega^s_{\epsilon} | \tilde{n}^s_{\epsilon},0) p(\beta^s_{\epsilon} | \mu_{\epsilon}) \right)
$$
where $k^s_{\epsilon} = \tilde{n}^s_{\epsilon 0} - \frac{\tilde{n}^s_{\epsilon}}{2}$

The MCMC sampling scheme is performed as follows.

\begin{itemize}

\item Update $\beta$

The posterior for (each) $\beta$ given $\omega$ is a normal as:

$$
p(\beta_{\epsilon}| \omega_{\epsilon}, \{ \tilde{n}_{\epsilon}\}) \propto
  e^{k_{\epsilon}\beta_{\epsilon}} e^{- \beta_{\epsilon}^2 \frac{\omega_{\epsilon}}{2}} 
  e^{\beta_{\epsilon} \frac{\mu_{\epsilon}}{\sigma_{\epsilon}^2} }  e^{- \frac{\beta_{\epsilon}^2}{2 \sigma_{\epsilon}^2}} =
$$
$$
e^{-\frac{\beta^2_{\epsilon}}{2} (\omega_{\epsilon} + \frac{1}{ \sigma_{\epsilon}^2}) + \beta_{\epsilon} (k_{\epsilon} + \frac{\mu_{\epsilon}}{\sigma_{\epsilon}^2})}
$$

Thus:

$$
\beta_{\epsilon}| \omega_{\epsilon}, \{ \tilde{n}_{\epsilon}\} \sim N \left( \left(k_{\epsilon} + \frac{\mu_{\epsilon}}{\sigma_{\epsilon}^2}\right) \frac{1}{\omega_{\epsilon} + \frac{1}{\sigma_{\epsilon}^2}}, \frac{1}{\omega_{\epsilon} + \frac{1}{\sigma_{\epsilon}^2}} \right)
$$

\item Update $\omega$

As in \citet{polson2013bayesian}, the posterior for (each) $\omega$ is:

$$
p(\omega_{\epsilon}| \beta_{\epsilon}, \{ \tilde{n}_{\epsilon}\}) \propto
e^{- \frac{\omega_{ij}\beta_{\epsilon}^2}{2}} p(\omega_{\epsilon} | \tilde{n}_{\epsilon},0) 
$$

Thus:

$$
\omega_{\epsilon}| (\beta_{\epsilon}, \{ \tilde{n}_{\epsilon}\} ) \sim \text{PG}(n_{\epsilon}, \beta_{\epsilon})
$$
that can be updated using the algorithm in \cite{polson2013bayesian}.

\item Update $\eta$

In order to sample $\eta$ more efficiently we integrate out $\beta$ from the posterior (we drop the subscripts for simplicity):

$$
p(\eta | \omega, \{ \tilde{n} \} ) = \int \prod_{s = 1,\dots,S} \prod_{\epsilon_1 \dots \epsilon_m} p(\eta | \omega, \beta, \{ \tilde{n} \} ) p(\beta) d\beta = p(\eta) \prod_{s = 1,\dots,S} \prod_{\epsilon_1 \dots \epsilon_m} \int e^{k \beta} e^{- \frac{\omega \beta^2}{2}} \frac{e^{-\frac{(\beta - \mu)^2}{2 \sigma^2}}}{\sqrt{2 \pi \sigma^2}} d\beta
$$

where:

$$
\int e^{k \beta} e^{- \frac{\omega \beta^2}{2}} \frac{e^{-\frac{(\beta - \mu)^2}{2 \sigma^2}}}{\sqrt{2 \pi \sigma^2}} d\beta = \frac{e^{-\frac{\mu}{2 \sigma^2}}}{\sqrt{2 \pi \sigma^2}} \int e^{- \frac{\beta^2}{2} (\omega + \frac{1}{ \sigma^2}) + \beta(k + \frac{\mu}{\sigma^2})}  d\beta \propto e^{-\frac{\mu}{2 \sigma^2}} e^{\frac{(k + \frac{\mu}{\sigma^2})^2}{2(\omega + \frac{1}{\sigma^2})}}
$$

$\quad$

However we can still write the posterior on $\eta$ given the variables $\beta$ as:

$$
\hspace{-1cm}
p(\eta, | \beta) \propto p(\eta)  \prod_{\epsilon_1 \dots \epsilon_m}  \prod_{s = 1}^S p(\beta^s_{\epsilon_1 \dots \epsilon_m} | \mu_{\epsilon_1 \dots \epsilon_m} )  \propto 
$$
$$
\underbrace{N(\mu | \mu_0, \Sigma_0) \text{Gamma}(\lambda_1 | a_{\lambda}, b_{\lambda}) \text{Gamma}(\lambda_2 | a_{\lambda}, b_{\lambda})}_{p(\eta)}   \prod_{\epsilon_1 \dots \epsilon_m}  \prod_{s = 1}^S N(\beta^s_{\epsilon_1 \dots \epsilon_m} | \mu_{\epsilon_1 \dots \epsilon_m}, (\sigma_{\epsilon_1 \dots \epsilon_m})^2 )
$$

\item Update $\{ n^s_{ij} \}$

These variables can be updated with a MH algorithm. The posterior can be written as

$$
p(n^s_{ij} | C_{k s}, \omega_{sij}, p_s) \propto p( n^s_{ij} | \omega_{ij}, \omega_s) \ p(C_{ks} | n^s_{ij}, p_s)
$$

where the first term is a Poisson distribution and the second term is a product of binomial distributions. 

\item Update $p_s$

The parameter can be sampled easily from the full conditional:

$$
p_S \sim \text{Beta}\left( 1 + \sum_{j=1}^K C_{js} , 1 + \sum_{j=1}^K N_{js} -\sum_{j=1}^K C_{js} \right)
$$

\end{itemize}

\subsection{Long count data model}
\label{mcmc3}

The parameters to be updated are the coefficient $\beta^{i,a}$ and $\beta^{i,d}$ assigning the masses to the entry and exit distribution for $1$-periods and $2$-periods, the means $\mu^a$ and $\mu^{d}$ of the prior, the stopping variables $S_{B_{\epsilon}}$, the number of latent individuals $n^{i_0}({i_1}, {j_1},{i_2},{j_2})$ and the detection probability $p$.

\begin{itemize}

\item Update $\beta_j^{i}$ (entry coefficients)

Each $\beta_j^{i}$ can be updated condition on the other parameters using the Polya-Gamma scheme for multinomial outcomes. We define as $n_i$ the sample size for $\beta_j^{i}$ (which is equal to $\#(\Omega_i)$) and as $a_{ij}$ the number of successes (equal to $\# (B^{i}_j)$).

Following \citet{polson2013bayesian}, the posterior of $\beta$ (we drop the index $i$ in the following) can be found as

$$
p(\beta_j | a_{j}, n ) \propto p(a_{j} | \beta_j, n) p(\beta_j) \propto \frac{(e^{\beta_j})^{a_{j}}}{(\sum_{k=1}^{K} e^{\beta_k})^{n}} e^{- \frac{(\beta_j - \beta^0_j)^2}{\sigma_0^2}} = 
$$
$$
\frac{(e^{\beta_j})^{a_{j}}}{\left( \underbrace{ \sum_{k=1 , k \ne j}^{K} e^{\beta_k}}_{e^{c_j}}  + e^{\beta_j} \right)^n} e^{- \frac{(\beta_j - \beta^0_j)^2}{\sigma_0^2}} \propto  \frac{(e^{\beta_j - c_j})^{a_{j}}}{(1 + e^{\beta_j - c})^n} e^{- \frac{(\beta_j - \beta^0_j)^2}{\sigma_0^2}} = 
$$
$$
2^{-n} e^{k(\beta_j - c_j)}  e^{- \frac{(\beta_j - \mu)^2}{\sigma^2}} \int e^{- \omega_j \frac{(\beta_j - c_j)^2}{2}} p(\omega_j) d\omega_j
$$

where $\omega_j \sim \text{PG}(n,0)$, $k = a_{j} - \frac{n}{2}$ and $c_j = \log(1 + \sum_{k \ne j} e^{\beta_k})$, from which it follows that a Gibbs sampler can be constructed as

$
\omega_j | \beta_j \sim \text{PG}(n, \beta_j - c_j) \\
\beta_j | \omega_j \sim N( (\frac{\sigma^2}{1 + \omega_j \sigma^2})^{-1} (k + \omega_j c_j + \frac{\beta^0_j}{\sigma_0^2}), \frac{\sigma_0^2}{1 + \omega_j \sigma_0^2} )
$

\item Update $\beta_{j,k}^{i}$ (exit coefficient)

If $n_{kj}$ is the number of elements in $B^{i}_{k,j}$, the likelihood of $\beta^i_{k,j}$ comes from assigning $n_{kj}$ elements to the set $B^{i}_{k,j}$ out of the $a_k$ elements in set $B^{i}_{k}$.

The posterior is (dropping the subscript $i$ and $k$)

$$
e^{- \frac{(\beta_j - \beta^0_j)^2}{\sigma_0^2}} \prod_{k=1}^j p(\beta_j | n_{ij}, a_i ) \propto e^{- \frac{(\beta_j - \beta^0_j)^2}{\sigma_0^2}} \prod_{k=1}^j \frac{(e^{\beta_j})^{n_{kj}}}{( \sum_{l=k}^{13} e^{\beta_l})^{a_k}} \propto
$$
$$ 
e^{- \frac{(\beta_j - \beta^0_j)^2}{\sigma_0^2}} \prod_{k=1}^j 2^{-a_k} e^{k_{kj}(\beta_j - c_{jk})} \int e^{- \omega \frac{(\beta_j - c_{jk})^2}{2}} p(\omega_k) d\omega_k
$$
 
where $\omega_k \sim \text{PG}(a_k,0)$, $k_{kj} = n_{kj} - \frac{a_k}{2}$ and $c_{ji} = \log(\sum_{k \ge i, k \ne j}^{K} e^{\beta_k})$. The posterior of $\beta_j$ can thus be expressed as

$$
e^{- \frac{(\beta_j - \beta^0_j)^2}{\sigma_0^2}} \prod_{k=1}^j e^{k_{kj}(\beta_j - c_{jk})} e^{- \omega_k \frac{(\beta_j - c_{jk})^2}{2}}
$$

from which it follows that a sample from the posterior can be performed by sampling

$$
\begin{cases}
\omega_k \sim PG(a_k, \beta_j - c_{jk}) \qquad k = 1,\dots,j\\
\beta_j | \omega_k \sim N( (\frac{\sigma_0^2}{1 + \sum_k \omega_k \sigma_0^2})^{-1} (\sum_k k_{kj} + \sum_k (\omega_k c_{jk}) + \frac{\beta^0_j}{\sigma_0^2}), \frac{\sigma_0^2}{1 + \sum_k \omega_k \sigma_0^2})
\end{cases}
$$

\item Update $\mu$

These parameters can be sampled straightforwardly from the full conditional, as the prior is $(\mu_1,\dots,\mu_{K_1}) \sim \text{GP}((\mu^0_1,\dots,\mu^0_{K_1}), \sigma^2_0 k_l)$ and the likelihood is $\beta^j_i \sim N(\mu_i, \sigma^2)$.

\item Update $\rho$

If $G$ is the joint entry and exit distribution and $S_{B_{\epsilon}}$ is the indicator variable which is $1$ if $B_{\epsilon}$ is not further partitioned and $0$ otherwise, we have

$$
G(x | x \in B_{\epsilon}) = S_{B_{\epsilon}} U_{B_{\epsilon}}(x) + (1 - S_{B_{\epsilon}}) \sum_{i=1}^{K_{\epsilon}} G( x | x \in B_{\epsilon, \epsilon_i}) 1(x \in B_{\epsilon, \epsilon_i})
$$


To sample each $S_{B_{\epsilon}}$, we first sample all the $S$ and $\beta$ for the ``children'' sets of $B$ and then sample $S_{B_{\epsilon}}$ from its posterior, which can be written as

$$
\hspace{-1cm}
p(S_{B_{\epsilon}} = 1 | \beta, x) =   p(S_{B_{\epsilon}} \frac{p(x | \beta, S_{B_{\epsilon}} = 1)}{p(x | \beta)} = 1) = \rho 
\frac{\text{U}_{B_i^1}(x)}{ \rho U_{B_{\epsilon}}(x) + (1 - \rho) \sum_{i=1}^{K_{\epsilon}} G( x | x \in B_{\epsilon, \epsilon_i}) 1(x \in B_{\epsilon, \epsilon_i}) } 
$$ 

\item Update $n_{d,w,m,y}$

We update this parameter with a MH step, where the posterior distribution is

$$
p(n^{i_0}({i_1}, {j_1},{i_2},{j_2}) | \omega^{i_0}({i_1}, {j_1},{i_2},{j_2}), \omega, \{ C^{i_0}_{i_1,i_2} \}, p) \propto  
$$
$$
p(n^{i_0}({i_1}, {j_1},{i_2},{j_2}) | \omega, \omega^{i_0}({i_1}, {j_1},{i_2},{j_2})) \ p( \{ C^{i_0}_{i_1,i_2} \} | \{ N^{i_0}_{i_1,i_2} \}, p)
$$

\item Update $p$

This parameter is sampled straightforwardly from its full conditional:

$$
p \sim \text{Beta}\left(1 + \sum_{i_0} \sum_{\epsilon} C^{i_0}_{\epsilon}, 
1 + \sum_{i_0} \sum_{\epsilon} N^{i_0}_{\epsilon} -  \sum_{i_0} \sum_{\epsilon} C^{i_0}_{\epsilon}\right)
$$

\end{itemize}

\bibliographystyle{Chicago}
\bibliography{biblio}